\pdfoutput=1
\documentclass{jfm}

\usepackage{graphicx}
\usepackage{epstopdf}
\usepackage{natbib}

\ifCUPmtlplainloaded \else
  \checkfont{eurm10}
  \iffontfound
    \IfFileExists{upmath.sty}
      {\typeout{^^JFound AMS Euler Roman fonts on the system,
                   using the 'upmath' package.^^J}%
       \usepackage{upmath}}
      {\typeout{^^JFound AMS Euler Roman fonts on the system, but you
                   dont seem to have the}%
       \typeout{'upmath' package installed. JFM.cls can take advantage
                 of these fonts,^^Jif you use 'upmath' package.^^J}%
      }
  \else
  \fi
\fi

\ifCUPmtlplainloaded \else
  \checkfont{msam10}
  \iffontfound
    \IfFileExists{amssymb.sty}
      {\typeout{^^JFound AMS Symbol fonts on the system, using the
                'amssymb' package.^^J}%
       \usepackage{amssymb}%
       \let\le=\leqslant  
         
      }{}
  \fi
\fi

\ifCUPmtlplainloaded \else
  \IfFileExists{amsbsy.sty}
    {\typeout{^^JFound the 'amsbsy' package on the system, using it.^^J}%
     \usepackage{amsbsy}}
    {\providecommand\boldsymbol[1]{\mbox{\boldmath $##1$}}}
\fi

\newsavebox{\astrutbox}
\sbox{\astrutbox}{\rule[-5pt]{0pt}{20pt}}

\usepackage{epstopdf}
\usepackage{wrapfig}
\usepackage{nomencl}
\usepackage{bm}
\usepackage{subcaption}
\usepackage[retainorgcmds]{IEEEtrantools}
\usepackage{natbib}
\usepackage{nomencl}
\usepackage[colorlinks]{hyperref}
\usepackage{amsmath}
\usepackage{bm}
\usepackage{layouts}
\usepackage[retainorgcmds]{IEEEtrantools}
\usepackage{natbib}

\newcommand{\ud}{\,\text{d}}
\renewcommand{\i}{\mathrm{i}}
\newcommand{\e}{\mathrm{e}}

\graphicspath{{Figs/}{IllustrationFigs/}{DataFigs/}}
\title{On the noise prediction for serrated leading edges}
\author[B. Lyu and M. Azarpeyvand]%
{B. Lyu$^1$
  \thanks{Email address for correspondence: bl362@cam.ac.uk}
and M. Azarpeyvand$^2$
}

\affiliation{$^1$Department of Engineering, University of Cambridge, Cambridge CB2 1PZ, UK\\[\affilskip]
$^2$Department of Mechanical Engineering, University of Bristol, Bristol BS8
1TR, UK\\[\affilskip]}

\pubyear{2016}
\volume{??}
\pagerange{???}
\date{?; revised ?; accepted ?. - To be entered by editorial office}

\hypersetup{citecolor = blue, linkcolor = blue}
\begin{document}
\maketitle
\begin{abstract} 
    An analytical model is developed for the prediction of noise radiated by an
    aerofoil with leading edge serration in a subsonic turbulent stream. The
    model makes use of the Fourier Expansion and Schwarzschild techniques in
    order to solve a set of coupled differential equations iteratively and
    express the far-field sound power spectral density in terms of the
    statistics of incoming turbulent upwash velocity. The model has shown that
    the primary noise reduction mechanism is due to the destructive
    interference of the scattered pressure induced by the leading edge
    serrations.  It has also shown that in order to achieve significant sound
    reduction, the serration must satisfy two geometrical criteria related to
    the serration sharpness and hydrodynamic properties of the turbulence. A
    parametric study has been carried out and it is shown that serrations can
    reduce the overall sound pressure level at most radiation angles,
    particularly at downstream angles close to the aerofoil surface. The sound
    directivity results have also shown that the use of leading edge serration
    does not particularity change the dipolar pattern of the far-field noise at
    low frequencies, but it changes the cardioid directivity pattern associated
    with radiation from straight-edge scattering at high frequencies to a
    tilted dipolar pattern. 
\end{abstract}

\section{Introduction}
The issue of noise generation from aerofoils has been the subject of much
theoretical, experimental and computational research over the past few decades
and is of great importance in many applications, such as jet engines, wind
turbine blades, high-speed propellers, helicopter blades, etc.  Aerofoil
noise can generally be categorised as self-noise and inflow-turbulence noise.
The aerofoil self-noise is due to the interaction of the aerofoil with its own
boundary layer and the flow instabilities present in the boundary
layer~\citep{Brooks1989}. The aerofoil inflow-turbulence noise, on the other
hand, is due to the interaction of an incoming unsteady gust with the aerofoil. The aerofoil inflow-turbulence interaction noise is a
significant contributor in systems involving multiple rows of blades, such as
jet engines and contra-rotating propellers. For instance, the wake flow shed by
the aircraft engine fan blades interacts with following blades and vanes,
causing leading edge noise from the rear blades. Likewise, the interaction of
the wake flow from the front row blades in a contra-rotating open rotor (CROR)
system with the downstream blades is considered as the main source of broadband
noise from such configurations. Also, the interaction of atmospheric turbulence
with the blades of wind turbines can also cause high level of
low-frequency broadband noise.

The prediction of leading-edge turbulence interaction noise has been the subject of much research over
the past decades~\citep{Sears1941, Graham1970, Amiet1975, Devenport2010}. Sears
originally considered the interaction of an unsteady sinusoidal gust with a
flat plate and developed a model for the prediction of the plate aerodynamic
response under such unsteady loading. Sears' model was later further developed
and extended to compressible flows by~\citet{Graham1970} and~\citet{Amiet1975}.
In Amiet's model, the blade response function to an incoming gust is first
obtained using the Schwarzschild technique and the far-field sound is then
formulated based on the theories of Kirchoff and~\citet{Curle1955} using the
radiation integral. Amiet's model shows that the far-field sound power spectral
density (PSD) is directly related to the energy spectrum of the velocity
fluctuations of the incoming gust.  It has been widely shown that Amiet's model
can provide fairly good comparisons with experimental observations when the
turbulence statistical quantities are known. The effects of aerofoil
geometrical parameters, such as angle of attack, aerofoil thickness, camber,
etc. on the generation of leading-edge turbulence interaction noise has also been the subject of some
theoretical works~\citep{Goldstein1976,Goldstein1978,Myers1995,Myers1997,
Atassi1993,Devenport2010,Roger2010}. 

The use of leading-edge treatments, inspired by the flippers of humpback whales
\citep{Bushnell1991,Fish1995,Miklosovic2004,Fish2008,Pedro2008} has been shown
to lead to improved aerodynamic and hydrodynamic performance, particularly at
high angles of attack. The recent extensive experimental works on the effects
of leading-edge serrations on the generation and control of turbulence
interaction noise has shown that the use of such treatments can result in
significant noise reduction over a wide range of
frequencies~\citep{Hansen2012,Narayanan2015}. For example, \citet{Narayanan2015}
showed that using sinusoidal leading-edge serrations for a flat plate and
NACA-65 type aerofoil leads to significant noise reduction.  Noise reductions
were found to be significantly higher for the flat plates. It was also shown
that the sound power reduction level is more sensitive to the serration
amplitude and less sensitive to the serration wavelength. In a more recent
study, it was shown that the use of complex leading-edge serrations, i.e.
serrations formed from the superposition of two serration profiles of different
frequency, amplitude and phase, can produce greater noise reductions than
single wavelength serrations~\citep{Chaitanya2016}. 

Besides the experimental activities, the problem of leading-edge noise
reduction using wavy edges has recently been studied in several computational
works~\citep{Lau2013,Kim2016,Turner2016}. In the work of \citet{Lau2013}, the
effectiveness of leading edge serrations for turbulence interaction noise
reduction was examined numerically. It was found that the hydrodynamic quantity
$k_1 h$ plays an important role in determining the effectiveness of the
serration, where $k_1$ is the hydrodynamic wavenumber of the disturbance in the
streamwise direction and the serration root-to-tip distance is $2h$. The
serration wavelength $\lambda$, on the other hand, was found to be less
important. However, one should note that the study assumed a perfect coherence
in the spanwise direction, which may not be the case in real-world
applications. The three-dimensionality of the disturbance was accounted for in
the work of~\citet{Kim2016} using synthetically generated turbulence. It was
argued that both a source cut-off and destructive interference effects
contributed to the sound reduction. Both of these two numerical works used a
regular sinusoidal serration profile attached to a flat plate. In a more recent
work by~\citet{Turner2016}, a dual-frequency wavy serration profile was
proposed and it was found that the more complex serration geometries can
increase the noise reduction. It is again worth noting that the upstream
disturbance was assumed to be perfectly correlated in the spanwise direction.

The above discussion provided a comprehensive literature review for the use of
leading edge serrations as a passive method for reduction of aerofoil
inflow-turbulence interaction noise. It is, however, worth mentioning that the
topic of leading edge serrations for improving the aerodynamic performance of
aerofoils has attracted much attention over the past few decades. A great deal
of experimental and numerical studies have been conducted to investigate the
effects of leading edge modifications on the aerofoil aerodynamic forces, early
separation and stall behaviour, unsteady forces, etc., at different flow
regimes. The detailed literature review of the aerodynamic performance of
serrated aerofoils is beyond the scope of the current study, but interested
readers are referred to several
works~\citep{Johari2007,Hansen2011,Miklosovic2004}.

Despite the significant body of work on noise reduction using leading-edge
serrations, no mathematical model has yet been developed to relate the radiated
noise to the serration geometrical parameters and turbulence quantities. While
the experimental observations~\citep{Paterson1976, Roger2010,
Roger2013,Narayanan2015} and computational works~\citep{Allampalli2009,
Atassi1993, Gill2013, Hixon2006,Lau2013,Kim2016,Turner2016} have provided the
evidence that leading-edge serrations can lead to significant noise reduction,
an accurate and robust analytical model can help better understand the
mechanism of such noise reductions. An accurate analytical model will also
enable us to assess the effectiveness of leading edge serrations at high Mach
numbers and Reynolds numbers, relevant to turbomachinery applications, where
numerical approaches are very costly and experiments difficult. This will also
provide us with a tool for blade design optimization purposes. In this paper,
we aim to extend Amiet's leading-edge noise prediction model to the case of
serrated leading-edges and provide a parametric study for the effects of
serrations on far-field noise. It will be shown in Section II that the
introduction of the serration will lead to a complex differential equation,
which is solved using the Schwarzschild technique in an iterative manner. The
scattered pressure loading will then be used in a radiation integral and the
far-field PSD will be found in terms of the incoming gust statistical
quantities and blade response function. Section III presents an extensive
parametric study of the proposed model and results will be provided for
far-field sound pressure level, noise directivity and overall sound pressure
level. The noise reduction mechanism will also be discussed in this section.
Section IV concludes the paper and lists our future plans.

\section{Analytical formulation} \label{sec:Formulation} 
In this section, we present a detailed derivation for the prediction of noise
due to the interaction of an unsteady gust with a flat plate with serrated
leading edge. The analytical model developed is based on Amiet's model and
Schwarzschild technique for solving the Helmholtz equation with appropriate
boundary conditions.  

\subsection{Leading edge noise modelling}
\label{sec:TheoreticalModelling}
\begin{figure} 
    \centering
    \includegraphics[width=0.6\textwidth]{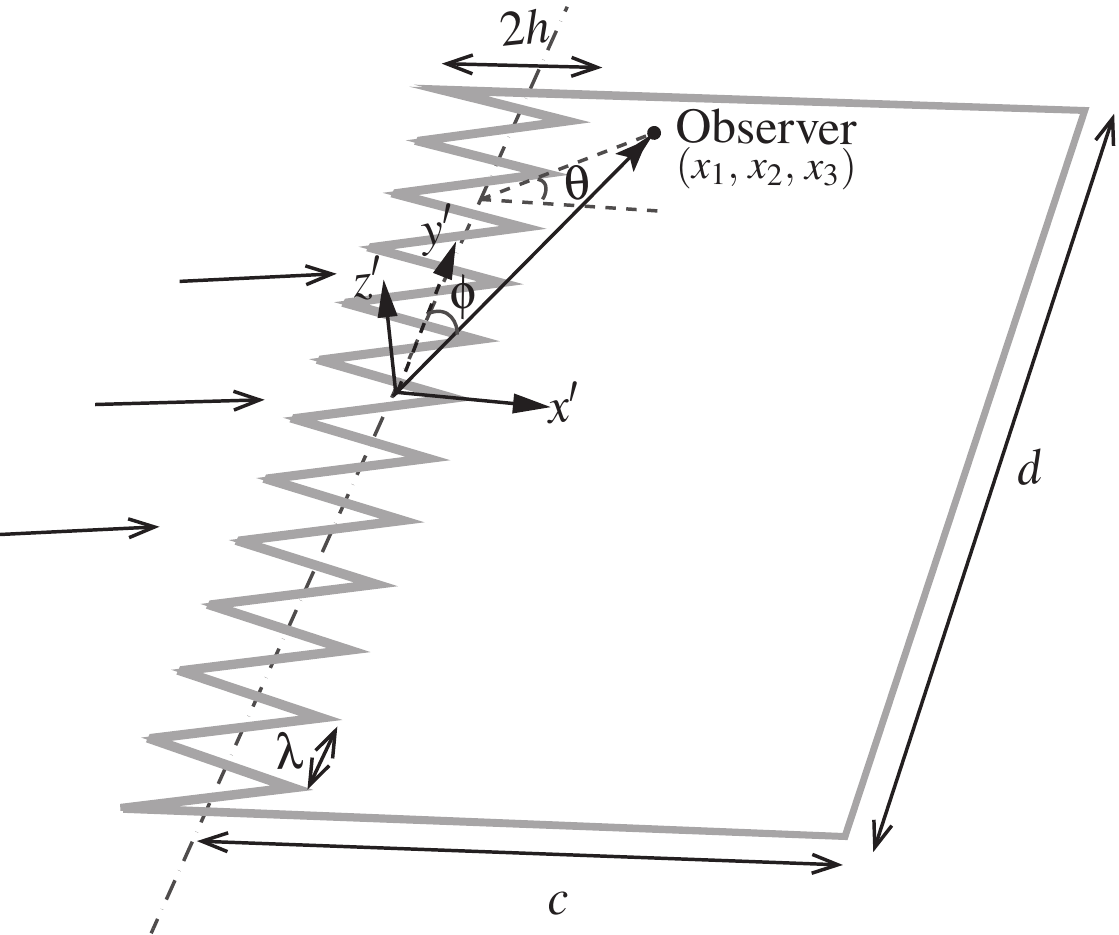}
    \caption{Schematic of a flat plate with a sawtooth-like leading edge.}
    \label{fig:SerratedFlatPlate} 
\end{figure} 

Let's consider an infinitesimally thin flat plate with leading edge serrations,
as shown in figure~\ref{fig:SerratedFlatPlate}, with an averaged chord length
$c$ and spanwise length $d$. A Cartesian coordinate system is chosen such that
the serration profile is an oscillatory function of zero mean.  When the
acoustic wavelength is smaller than the chord length $c$, the flat plate can be
considered as an infinitely long plate without a trailing
edge~\citep{Amiet1976, Amiet1978, Roger2005}. When the frequencies are high
enough as to the semi-infinite simplification is permissible, the plate can
also considered infinite in the spanwise direction when it has a relatively
large aspect ratio, typically larger than 3~\citep{Amiet1978, Roger2010}. Let
$x^{\prime}$, $y^{\prime}$ and $z^{\prime}$ denote the streamwise, spanwise and
normal directions to the plate, respectively. The observer point is located at
$(x_1,x_2,x_3)$, as shown in figure~\ref{fig:SerratedFlatPlate}.

\nomenclature[a-x]{$x^{\prime}$}{Chordwise axis in aerofoil-fixed frame}
\nomenclature[a-y]{$y^{\prime}$}{Spanwise axis in aerofoil-fixed frame}
\nomenclature[a-z]{$z^{\prime}$}{Axis perpendicular to aerofoil in
aerofoil-fixed frame} \nomenclature[a-c]{$c$}{Averaged chord length}
\nomenclature[a-d]{$d$}{Span length} \nomenclature[a-x1]{$x_1$}{The projection
    of observer point on $x^{\prime}$} \nomenclature[a-x2]{$x_2$}{The
	projection of observer point on $y^{\prime}$}
	\nomenclature[a-x3]{$x_3$}{The projection of observer point on
	    $z^{\prime}$}

As mentioned above, the origin of the coordinate system ($x^\prime, y^\prime,
z^\prime$) is chosen such  that
the serration profile, $H(y^\prime)$, is an oscillatory function of zero mean
and that $H(y^\prime)= 0$ in the absence of serrations.  Though the method
developed in this section can be used for any general periodic serrations, in
this paper we only focus on sawtooth serration, as shown in
figure~\ref{fig:SerratedFlatPlate}, where the root-to-tip length is $2h$ and
the serration wavelength is $\lambda$.
\begin{wrapfigure}{R}{0.35\textwidth}
    \includegraphics[width=\linewidth]{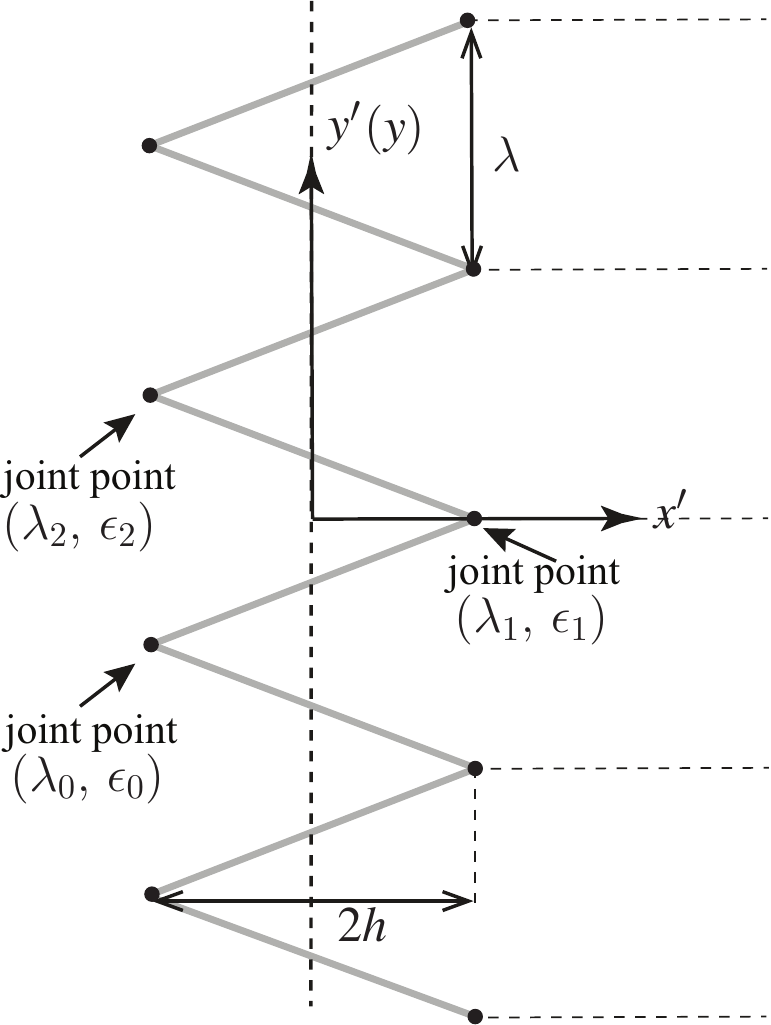} \caption{The
    schematic of serration profile.} \label{fig:serrationProfile}
\end{wrapfigure} 
The parameter $\sigma=4h/\lambda$ will
also be used to quantify the sharpness of the sawtooth serrations. To obtain a
mathematical description of $H(y^\prime)$, let us consider a single sawtooth
centred around the coordinate origin, and let $(\lambda_0, \epsilon_0)$,
$(\lambda_1, \epsilon_1)$ and $(\lambda_2, \epsilon_2)$ denote the three
joint-points defining this single sawtooth, as shown in
figure~\ref{fig:serrationProfile}. The serration profile function $H(y^\prime)$
can therefore be defined as 
\begin{equation} H(y^\prime) =
    \begin{cases} \sigma_0(y^\prime - \lambda_0 - m\lambda) + \epsilon_0,  &
	\lambda_0+m\lambda < y^\prime \le \lambda_{1}+m\lambda\\
	\sigma_1(y^\prime - \lambda_1 - m\lambda) + \epsilon_1,  &
	\lambda_1+m\lambda < y^\prime \le \lambda_{2}+m\lambda\\ \end{cases},
    \label{equ:ProfileFunction}
\end{equation} 
where $\sigma_j = (\epsilon_{j+1} - \epsilon_j)/(\lambda_{j+1} - \lambda_j)$,
$j = 0, 1$ and $m = 0$, $\pm1, \pm2, \pm3\cdots$. 
\nomenclature[a-H]{$H$}{Serration profile function} 
\nomenclature[a-h]{$h$}{Half of root-to-tip length of serrations}
\nomenclature[g-lambda]{$\lambda$}{Wavelength of serrations}

In this paper, we focus our attention on the unsteady upwash
disturbance~\citep{Amiet1975}, denoted by $w$, that exists upstream of the
leading edge, convecting downstream with the mean flow at the speed of $U$.
According to \citet{Kovasznay1953}, the unsteady motion on a uniform flow can
be decomposed into vorticity, entropy and sound-wave modes.  When perturbation
amplitudes are small as to linearisation is permissible, the three modes are
mutually independent. It is generally accepted that the incident turbulence can
be well presented by vorticity mode, which convects at speed of the mean flow.
Therefore, this study follows the same simplification and the incoming gust is
assumed to be frozen in the frame moving with the mean flow, i.e. the velocity
distribution is $w(x^\prime, y^\prime, t) = w_m(x^\prime - Ut, y^\prime)$,
where $t$ denotes time, for some function $w_m(x_m, y_m)$ describing the
distribution of upwash velocity in the travelling coordinate system $\{x_m,
y_m, z_m\}$. In this paper we make use of the same assumption. In the plate fixed
frame, the incoming gust can be written in terms of its wavenumber components,
$\tilde{w}(k_1, k_2)$, as 
\begin{equation} 
    w(x^\prime, y^\prime, t)  = \iint_{-\infty}^{\infty}
    \tilde{w}(k_1, k_2) \e^{\i (k_1(x^\prime - U t) + k_2 y^\prime)} \ud k_1 \ud
    k_2,  \label{equ:wTime} 
\end{equation} 
where the Fourier components $\tilde{w}(k_1, k_2)$ is given by 
\begin{equation} 
    \tilde{w}(k_1, k_2) =
    \left(\frac{1}{2\pi}\right)^2 \iint_{-L}^{L} w_m(x_m, y_m) \e^{-\i(k_1 x_m +
    k_2 y_m)} \ud x_m \ud y_m, \label{equ:FT} 
\end{equation} 
where $L$ is a large but finite number to avoid convergence difficulties and
$k_1$ and $k_2$ denote the Fourier wavenumbers in the streamwise and spanwise
directions, respectively.  Using (\ref{equ:wTime}) one can find
\begin{IEEEeqnarray}{rCl} 
    w(x^\prime, y^\prime, \omega) &=& \frac{1}{2\pi} \int
    w(x^\prime, y^\prime, t) \e^{\i\omega t} \ud t \nonumber\\ & = & \frac{1}{U}
    \int_{-\infty}^{\infty} \tilde{w}(\omega/ U, k_2) \e^{\i(\omega x^\prime / U+
    k_2 y^\prime)} \ud k_2, \label{equ:generalGust} 
\end{IEEEeqnarray} 
where $\omega$ represents angular frequency.

Equation~(\ref{equ:generalGust}) suggests that the general unsteady gust can be
decomposed into a set of plane-wave-like gusts, each of which taking the form
of
\begin{equation} w_i = w_{ia}\e^{-\i(\omega t-k_1 x^{\prime}-k_2 y^{\prime})},
    \label{equ:GustForSerrated} 
\end{equation}
where, $k_1 = \omega / U$ and $w_{ia}$ denotes the magnitude of the
upwash velocity and is a function of both $\omega$ and $k_2$. The scattered
velocity potential $\phi_t$ is governed by the convective wave equation, i.e.  
\begin{equation} \nabla^2 \phi_t -
    \frac{1}{c_0^2} \left(\frac{\partial}{\partial t} + U
    \frac{\partial}{\partial x^{\prime}}\right)^2 \phi_t = 0,
    \label{equ:TotalEquation} 
\end{equation} 
where $c_0$ denotes the speed of sound. Hence, if we can find the far-field
sound induced by a single gust by solving the wave equation subject to
appropriate boundary condition upstream of the leading edge and over the
surface of the serrated plate, then a more general solution can readily be
obtained by performing an integration over $k_2$, as shown in
(\ref{equ:generalGust}). The next part of this section is therefore devoted to
the single-gust solution for a flat plate with leading-edge serration. 

\nomenclature[a-i]{$i$}{Imaginary symbol $\sqrt{-1}$}
\nomenclature[a-w_i]{$w_i$}{Upwash velocity of upstream gust}
\nomenclature[a-phi]{$\phi$}{Second-part scattered velocity potential field}
\nomenclature[g-o]{$\omega$}{Angular frequency $\omega = 2\pi f$}
\nomenclature[a-k_1]{$k_1$}{Hydrodynamic wavenumber in chordwise direction}
\nomenclature[a-k_2]{$k_2$}{Hydrodynamic wavenumber in spanwise direction}
\nomenclature[a-t]{$t$}{Time} \nomenclature[a-Uc]{$U_c$}{Turbulence convection
velocity} \nomenclature[a-Phiia]{$\Phi_{ia}$}{Magnitude of initial potential
solution}

\subsection{Single-gust solution}
The full solution $\phi_t$ to (\ref{equ:TotalEquation}) can be written in terms
of an initial and a residual potential part. The initial potential, $\phi_i$,
is used to cancel the upwash velocity on the plane $z^\prime = 0$. Upon
defining $k = \omega/c_0$, $\beta^2 = 1-M_0^2$ and $M_0 = U/c_0$, one can show
that on the plane $z^\prime = 0$, $\phi_i$ takes the form of
\begin{equation}
  \phi_i = - \Phi_{ia}\e^{-\i(\omega t-k_1 x^{\prime}-k_2 y^{\prime})},
  \label{equ:initialPotential}
\end{equation}
where $\Phi_{ia} \equiv -{w_{ia}}/\sqrt{(k_1 \beta + k M_0/\beta)^2 + k_2^2 -
(k / \beta)^2 }$. 

The Schwarzschild technique can then be used to calculate the
second part, i.e. residual term, of the potential field, $\phi$, which would
cancel the potential field of the initial solution upstream of the leading edge
(such that we have $\phi_t = 0$ for $x^\prime < 0$ and $z^\prime = 0$). Thus,
the boundary conditions at $z^{\prime} = 0$ for $\phi$ read
\begin{equation}
  \left\{
    \begin{aligned}
      &\frac{\partial \phi}{\partial z^{\prime}} = 0, &x^{\prime} > H(y^{\prime})\\
      & \phi = \Phi_{ia}\e^{-\i(\omega t - k_1x^{\prime} -k_2 y^{\prime})}, & x^{\prime} \le H(y{^\prime}).
    \end{aligned}
  \right.
  \label{equ:BoundaryConditionsForSerrated}
\end{equation}
The equation governing the second-part potential field $\phi$ remains unchanged, i.e.
\begin{equation}
  \nabla^2 \phi - \frac{1}{c_0^2} \left(\frac{\partial}{\partial t} + U
  \frac{\partial}{\partial x^{\prime}}\right)^2 \phi = 0.
  \label{equ:TimeWaveEquation}
\end{equation}
Equation~(\ref{equ:TimeWaveEquation}) together with the boundary conditions
given in (\ref{equ:BoundaryConditionsForSerrated}) forms a well-posed
mathematical problem, and we attempt to solve it in this section.

\subsubsection{Boundary-value problem} 
With the assumption of harmonic perturbation $\phi =
\Phi(x^{\prime},y^{\prime},z^{\prime})\e^{-\i \omega t}$,
(\ref{equ:TimeWaveEquation}) reduces to
\begin{equation}
  \beta^2 \frac{\partial^2 \Phi}{\partial x^{\prime 2}} + \frac{\partial^2
  \Phi}{\partial z^{\prime 2}} +\frac{\partial^2 \Phi}{\partial y^{\prime 2}} +
  2\i kM_0\frac{\partial \Phi}{\partial x^{\prime}} + k^2 \Phi = 0.
\end{equation}

\nomenclature[a-u]{$U$}{Uniform flow velocity}
\nomenclature[a-Phi]{$\Phi$}{Time Fourier transformation of second-part scattered potential field}
\nomenclature[a-c]{$c_0$}{Speed of sound}
\nomenclature[a-k]{$k$}{Acoustic wavenumber $\omega/c_0$}
\nomenclature[g-beta]{$\beta$}{$\beta = \sqrt{1-M_0^2}$}
\nomenclature[a-M0]{$M_0$}{Mach number of uniform flow}

In order to make the boundary conditions in
(\ref{equ:BoundaryConditionsForSerrated}) independent of $y^{\prime}$, the
coordinate transformation $x = x^{\prime} - H(y^{\prime}), y  = y^{\prime}, z =
z^{\prime}$ is used~\citep{Roger2013}, which leads to the following
differential equation~\citep{Sinayoko2014,Lyu2015,Lyu2016b}:
\begin{equation}
  \left(\beta^2+{H^\prime}^2(y)\right)\frac{\partial^2 \Phi}{\partial x^2} +
  \frac{\partial^2 \Phi}{\partial y^2} + \frac{\partial^2 \Phi}{\partial z^2} -
  2H^{\prime}(y)\frac{\partial^2 \Phi}{\partial x\partial y} + \left(2\i M_0k-H^{\prime \prime}(y)\right)\frac{\partial \Phi}{\partial x} + k^2\Phi = 0,
  \label{equ:StretchedWaveEquation}
\end{equation}
where $H^{\prime}(y)$ and $H^{\prime \prime}(y)$ denote the first and second
derivatives of $H(y)$ with respect to $y$. The boundary conditions now read
\begin{equation}
  \begin{cases}
    \Phi(x, y, 0) =  \Phi_{ia}\e^{\i(k_1 x + k_2 y)} \e^{\i k_1 H(y)}, & x \le 0 \\
    \partial \Phi(x, y, 0)/\partial z= 0, & x > 0. \\
  \end{cases}
  \label{equ:StretchedBoundaryConditions}
\end{equation}

The set of equations~(\ref{equ:StretchedWaveEquation})
and~(\ref{equ:StretchedBoundaryConditions}) forms a linear boundary-value
problem. However, unlike the governing equation for a straight leading
edge~\citep{Amiet1976b}, the coefficients in (\ref{equ:StretchedWaveEquation})
now depend on $y$, and therefore the standard ``separation of variables''
technique cannot be easily applied. Therefore, a Fourier expansion technique
will be initially employed to eliminate the $y$ dependency in
(\ref{equ:StretchedWaveEquation}), as explained in the following section.
\nomenclature[a-x]{$x$}{Chordwise axis in stretched coordinate system}
\nomenclature[a-y]{$y$}{Spanwise axis in stretched coordinate system}
\nomenclature[a-z]{$z$}{Axis perpendicular to aerofoil in stretched coordinate
system}

\subsubsection{Fourier expansion} \label{sec:FloquetAndFourier} 
Using both the infinite-span and serration periodicity assumptions, one can
make use of the Fourier series in terms of the new coordinates ($x, y, z$) to
expand the induced potential due to the gust interaction , as
\begin{equation}
  \Phi(x,y,z) = \sum_{-\infty}^{\infty}\Phi_n(x,z)\e^{\i k_{2n}y},
  \label{expansion}
\end{equation}
where $k_{2n} = k_2 + 2n\pi/\lambda$.
\nomenclature[a-Phin]{$\Phi_n$}{Second-part scattered potential of mode $n$}
\nomenclature[a-k2n]{$k_{2n}$}{Characteristic wavenumber of mode $n$, $k_{2n} =
k_2 + 2 n \pi/\lambda$} \nomenclature[t-']{$\prime$}{First derivative of
function} \nomenclature[t-'']{$\prime\prime$}{Second derivative of function}
Substituting this expansion into (\ref{equ:StretchedWaveEquation}) and
multiplying the resulted equation by $\e^{-\i k_{2n^{\prime}}y}$, then integrating
over $y$ from $-\lambda/2$ to $\lambda/2$, one can readily show that
\begin{equation}
  \begin{aligned}
    \left\{	\beta^2\frac{\partial^2}{\partial x^2} \right. +
    &\left.\frac{\partial^2 }{\partial z^2} + 2\i kM_0\frac{\partial }{\partial x} +  (k^2 - k_{2n^{\prime}}^2) \right\}\Phi_{n^{\prime}}\\
    & + \frac{1}{\lambda}\int_{-\lambda/2}^{\lambda/2} \sum_{n =
    -\infty}^{\infty} \left\{  H^{\prime 2} \frac{\partial^2}{\partial x^2}
    -(H^{\prime \prime}+2\i k_{2n} H^{\prime})\frac{\partial}{\partial x}
\right\}\Phi_n \e^{\i 2(n-n^{\prime})\pi/\lambda y} \ud y=0.
    \label{equ:GeneralCoupledEquation}
  \end{aligned}
\end{equation}

If both $H^{\prime}(y)$ and $H^{\prime \prime}(y)$ were constant within the
entire wavelength, the summation over different modes in
(\ref{equ:GeneralCoupledEquation}) would vanish and one would obtain an
equation which only involves one mode, say $n^{\prime}$. However, for the
profile of the sawtooth serration, $H^{\prime}(y)$, is not continuous and hence
$H^{\prime \prime}(y)$ is singular at the joint-points
($\lambda_i,\epsilon_{i}$). We use the generalized function $\delta(y)$ to
describe the singularities, i.e.
\begin{equation}
  \begin{aligned}
    &	H^{\prime}(y) =
    \begin{cases}
      \sigma_0 ,  & \lambda_0+m\lambda < y \le \lambda_{1}+m\lambda\\
      \sigma_1,  & \lambda_1+m\lambda < y \le \lambda_{2}+m\lambda\\
    \end{cases},\\
    &H^{\prime \prime}(y) = \sum_{m = -\infty}^{\infty} (-1)^{m+1}2\sigma \delta(x-m\lambda/2).
  \end{aligned}
  \label{equ:1stDerivative}
\end{equation}
where $\sigma = 4 h / \lambda$ signifies the serration sharpness. As
$\int_{-\infty}^{\infty} \delta(x)f(x) \ud x = f(0)$, the summation in
(\ref{equ:GeneralCoupledEquation}) cannot be dropped, indicating that different
modes are coupled together. Substituting the serration profile function and its
derivatives, (\ref{equ:ProfileFunction}) and~(\ref{equ:1stDerivative}), into
(\ref{equ:GeneralCoupledEquation}), and making use of the fact that
$\int_{-\infty}^{\infty} f(x) \delta(x-\tau) \ud x = f(\tau)$, we obtain
\begin{equation}
  \begin{aligned}
    \left\{\left(\beta^2+\sigma^2\right) \frac{\partial^2}{\partial x^2} +
    \frac{\partial^2}{\partial z^2} + 2\i kM_0 \frac{\partial}{\partial x} + (k^2-k_{2n^{\prime}}^2)\right\} \Phi_{n^{\prime}}  \\
    = - \frac{4\sigma}{\lambda}\sum_{n-n^{\prime} = odd} \left(1-\frac{k_2\lambda + 2n\pi}{(n-n^{\prime})\pi}\right)\frac{\partial \Phi_{n}}{\partial x}.
  \end{aligned}
  \label{equ:coupled}
\end{equation}

\nomenclature[g-delta]{$\delta$}{Generalized function defined in (\ref{equ:DeltaDefinition})}
\nomenclature[g-simga]{$\sigma$}{Absolute value of the slope of a sawtooth edge}
\nomenclature[b-j]{$j$}{The $j$-th edge of a single piece of sawtooth serrations}
\nomenclature[g-lambdaj]{$\lambda_j$}{The spanwise coordinate of the starting point of the $j$-th edge }
\nomenclature[g-epsilonj]{$\epsilon_j$}{The chordwise coordinate of the starting point of the $j$-th edge}
\nomenclature[g-lambdajp]{$\lambda_{j+1}$}{The spanwise coordinate of the ending point of the $j$-th edge }
\nomenclature[g-epsilonjp]{$\epsilon_{j+1}$}{The chordwise coordinate of the ending point of the $j$-th edge}
\nomenclature[g-sigmaj]{$\sigma_j$}{Slope of the $j$-th edge }

We can write the set of differential equations obtained above in a more compact matrix form. Using a linear operator
\begin{equation}
  \mathcal{D} = \left\{\left(\beta^2+\sigma^2\right) \frac{\partial^2}{\partial
  x^2} + \frac{\partial^2}{\partial z^2} + 2\i kM_0 \frac{\partial}{\partial x}  \right\},
  \label{equ:Operator}
\end{equation}
and a vector of functions
\begin{equation}
  \boldsymbol{\Phi} = \left(
  \cdots \Phi_{-n^{\prime}}(x,z)\text{, }
  \Phi_{-n^{\prime}+1}(x,z)\text{, }\cdots
  \Phi_{n^{\prime}-1}(x,z)\text{, }
  \Phi_{n^{\prime}}(x,z)\text{, }\cdots
  \right)^\mathrm{T},
  \label{equ:VectorP}
\end{equation}
the coupled equations in  (\ref{equ:coupled}) can be written as
\begin{equation}
  \mathcal{D}\boldsymbol{\Phi} = \boldsymbol{A} \boldsymbol{\Phi} + \boldsymbol{B}\frac{\partial \boldsymbol{\Phi}}{\partial x},
  \label{equ:MatrixFormWaveEquations}
\end{equation}
where the symbol $\mathrm{T}$ in (\ref{equ:VectorP}) denotes the transpose of a
matrix. Matrices $\boldsymbol{A}$ and $\boldsymbol{B}$ denote the coefficient
matrices of $\boldsymbol{\Phi}$ and $\partial \boldsymbol{\Phi}/\partial x$,
respectively, and the elements $A_{ml}$ and $B_{ml}$, representing the entry
corresponding to mode $m$ in row and $l$ in column of matrix $\boldsymbol{A}$
and $\boldsymbol{B}$, are given by
\begin{equation}
  \begin{aligned}
    &A_{ml} = (k_{2m}^2 - k^2)\delta_{ml},
    &B_{ml} = \Bigg\{
      \begin{aligned}
        &\frac{4\sigma}{\lambda}\frac{m+l+k_2\lambda/\pi}{l-m}, &m-l = odd\\
        &0, &m-l = even,
      \end{aligned}
    \end{aligned}
    \label{equ:ABDefinition}
  \end{equation}
  where $\delta_{ml}$ represents the Kronecker delta.

  \nomenclature[a-D]{$\mathcal{D}$}{Linear differential operator}
  \nomenclature[a-Phi]{$\boldsymbol{\Phi}$}{Vector expression of the second-part scattered potential of different modes}
  \nomenclature[a-A]{$\boldsymbol{A}$}{Coefficient matrix for $\boldsymbol{\Phi}$ defined in (\ref{equ:MatrixFormWaveEquations})}
  \nomenclature[a-B]{$\boldsymbol{B}$}{Coefficient matrix for $\partial \boldsymbol{\Phi}/\partial x$ defined in (\ref{equ:MatrixFormWaveEquations})}
  \nomenclature[a-A]{$A_{ml}$}{Elements of matrix $\boldsymbol{A}$}
  \nomenclature[a-B]{$B_{ml}$}{Elements of matrix $\boldsymbol{B}$}
  \nomenclature[g-detla]{$\delta_{ml}$}{The Kronecker delta}
  \nomenclature[g-tau]{$\tau$}{Variable in function $\delta(x-\tau)$}
  \nomenclature[a-T]{$\mathrm{T}$}{Matrix transpose}

The boundary condition for each mode $n$ can be obtained by substituting the
profile geometry, (\ref{equ:ProfileFunction}), into the boundary conditions,
(\ref{equ:StretchedBoundaryConditions}), and performing the same Fourier
expansions,
  \begin{equation}
    \left\{
      \begin{aligned}
        &\Phi_n(x,0) = \Phi_{ia}a_n\e^{\i k_1x} ,& x \le 0\\
        &\frac{\partial \Phi_n}{\partial z}(x,0) = 0, &x > 0,
      \end{aligned}
      \right.
      \label{equ:MatrixFormBoundaryConditions}
  \end{equation}
  where $a_n$ is defined as
  \begin{equation}
    a_n = \frac{1}{\lambda}\int_{-\lambda/2}^{\lambda/2} \e^{\i k_1 H(y)} \e^{-\i 2n\pi/\lambda y} \ud y.
    \label{an}
  \end{equation}

  \nomenclature[a-an]{$a_n$}{Fourier coefficient of mode $n$}

Before we attempt to solve (\ref{equ:MatrixFormWaveEquations}), it is worth
examining some of its important properties. The matrix $\boldsymbol{A}$ is
obviously a diagonal matrix and if $\boldsymbol{B}$ was also diagonal, we would
be able to solve each mode individually, i.e. no mode coupling.  However,
$\boldsymbol{B}$ is not a diagonal matrix and different modes are coupled
together, in the sense that $\Phi_{n}$ for example, appears in the governing
equation of $\Phi_{m}$.  This means that every mode is interacting with the
other modes and cannot be solved individually. Also, it can be observed from
the expression of $\boldsymbol{B}$ in (\ref{equ:ABDefinition}) that the
strength of the modes coupling is proportional to $\sigma/\lambda$. This
indicates that the mode-coupling becomes stronger for sharper serrations.
Obviously, the mode-coupling phenomenon fades away when the serration amplitude
$2h$ is very small and the solution reduces to that of Amiet's for a straight
leading-edge. Also, at very low frequencies, it is expected that the
contribution of higher order modes becomes negligible compared with the 0-th
mode. Thus, for the equation governing the 0-th mode, the coupling with higher
modes becomes weak. One can, therefore, solve the 0-th mode individually, and
calculate the potential with only the contribution of the 0-th mode. But this
only works for very low frequencies or for very wide serrations. The coupling
effect becomes more pronounced at high frequencies. To solve these coupled
equations at relatively high frequencies, we will use an iterative procedure,
which is explained in the following section.

  \nomenclature[a-Phii]{$\boldsymbol{\Phi}^{(i)}$}{The $i$-th iterated solutions of PDEs}

  \subsubsection{Induced potential field} \label{sec:ExtensionOfAmiet}
  To obtain the induced potential field, (\ref{equ:MatrixFormWaveEquations})
  together with the boundary conditions in
  (\ref{equ:MatrixFormBoundaryConditions}) needs to be solved. Recall a set of
  linear algebraic equations, it is known that its solution can be sought via
  the so-called iterative process~\citep{Suli2003}. One can draw an analogy
  between these PDEs and the linear algebraic equations. In what follows, we
  shall explain the iterative procedure employed for solving our set of
  PDEs~\citep{Lyu2015, Lyu2016}.

  Substituting an assumed initial value $\boldsymbol{\Phi}^{(0)}$ into the
  coupling term in (\ref{equ:MatrixFormWaveEquations}), one can obtain
  \begin{equation}
    \mathcal{D}\boldsymbol{\Phi} = \boldsymbol{A} \boldsymbol{\Phi} + \boldsymbol{B}\frac{\partial \boldsymbol{\Phi}^{(0)}}{\partial x}.
    \label{equ:iteration1st}
  \end{equation}
  Solving (\ref{equ:iteration1st}) gives a new set of solutions
  $\boldsymbol{\Phi}^{(1)}$. Replacing $\boldsymbol{\Phi}^{(0)}$ in
  (\ref{equ:iteration1st}) with $\boldsymbol{\Phi}^{(1)}$, we obtain a new wave
  equation,
  \begin{equation}
    \mathcal{D}\boldsymbol{\Phi} = \boldsymbol{A} \boldsymbol{\Phi} + \boldsymbol{B}\frac{\partial \boldsymbol{\Phi}^{(1)}}{\partial x}.
    \label{equ:iteration2nd}
  \end{equation}
  Again, solving (\ref{equ:iteration2nd}) gives a new set of solutions
  $\boldsymbol{\Phi}^{(2)}$. Continuing this process, we obtain a solution
  sequence, $\boldsymbol{\Phi}^{(0)}$, $\boldsymbol{\Phi}^{(1)}$,
  $\boldsymbol{\Phi}^{(2)}$, $\boldsymbol{\Phi}^{(3)}\cdots$. If the sequence
  appears to be convergent, we manage to obtain the solution to
  (\ref{equ:MatrixFormWaveEquations}).

  The initial value $\boldsymbol{\Phi}^{(0)}$ used to start the first iteration
  can be found by ignoring all the coupling terms, i.e. with
  $\boldsymbol{B} = 0$, and by solving each equation individually using the
  standard Schwarzschild technique. The solution to each equation in the
  uncoupled matrix equation
  \begin{equation}
    \mathcal{D}\boldsymbol{\Phi} = \boldsymbol{A} \boldsymbol{\Phi}
    \label{initial}
  \end{equation}
  can be found as follows. Upon making use of the transformation of
  \begin{equation}
  \Phi_{n^\prime} = \bar{\Phi}_{n^\prime} \e^{-\i kM_0/(\beta^2 + \sigma^2)x},
  \end{equation}
  the individual equations in (\ref{initial}) reduce to
  \begin{equation}
    \left\{(\beta^2 + \sigma^2) \frac{\partial^2}{\partial x^2} +
    \frac{\partial^2}{\partial z^2} + K_{n^\prime}^2(\beta^2 + \sigma^2)\right\} \bar{\Phi}_{n^\prime} = 0,
    \label{equ:initialEle}
  \end{equation}
  where
  \begin{equation}
    K_{n^{\prime}} = \sqrt{k^2(1+\sigma^2) - k_{2n^{\prime}}^2(\beta^2+\sigma^2)}/(\beta^2+\sigma^2).
    \label{equ:Kndefinition}
  \end{equation}
  Making use of $X = x$ and $Z = \sqrt{\beta^2 + \sigma^2}z$ transformations,
  one can verify that (\ref{equ:initialEle}) reduces to the standard
  Schwarzschild problem, as
  \begin{equation}
    \left\{
      \begin{aligned}
        &\left( \frac{\partial}{\partial X^2} + \frac{\partial}{\partial Z^2} + K_{n^\prime}^2 \right) \bar{\Phi}_{n^\prime}(X, Z) = 0, \\
        &\bar{\Phi}_{n^\prime}(X, 0) = \Phi_{ia} a_{n^\prime} \e^{\i k_1 X}\e^{\i\frac{kM_0}{\beta^2+\sigma^2}X}, \quad X \le 0,  \\
        &\frac{\partial \bar{\Phi}_{n^\prime}}{\partial X}(X, 0) = 0, \quad X > 0.
      \end{aligned}
    \right.
    \label{equ:SchwartzProblem}
  \end{equation}
  The Schwarzschild technique entails that, for $X > 0$, i.e. over the
  plate, the solution for (\ref{equ:SchwartzProblem}) can be found from~\citep{Landahl1961,Amiet1976,Amiet1978} 
\begin{equation}
    \bar{\Phi}_{n^\prime}(X, 0) = \frac{1}{\pi} \int_{-\infty}^{0}
    \sqrt{\frac{-X}{\xi}} \frac{\e^{\i K_{n^\prime}(X-\xi)}}{X-\xi}
    \bar{\Phi}_{n^\prime}(\xi, 0) \ud\xi.
  \label{SchwarzschildSolution}
\end{equation}
Evaluating (\ref{SchwarzschildSolution}) and transforming back to the physical coordinate system ($x$) yields the initial solution 
  \begin{equation}
    \Phi^{(0)}_{n^{\prime}} = - \Phi_{ia} \e^{\i k_1 x} a_{n^{\prime}}
    \big((1+\i)E^\ast(\mu_{n^{\prime}} x) - 1\big),
    \label{equ:initialResults}
  \end{equation}
  where $\Phi^{(0)}_{n^{\prime}}$ is the element of vector $\boldsymbol{\Phi}^{(0)}$ corresponding to the $n^{\prime}$-th mode, and
  \begin{equation}
    \begin{aligned}
      &\mu_{n^{\prime}} = -K_{n^{\prime}} + k_1 +\frac{kM_0}{\beta^2 + \sigma^2},\\
      &E^\ast(x) = \int_0^x \frac{\e^{-\i t}}{\sqrt{2\pi t}} \ud t.
    \end{aligned}
  \end{equation}
  The initial solutions obtained by ignoring all the coupling terms denote the
  non-coupled part of the exact solution of each mode, which implies that an
  $n$-th mode excitation ($x < 0$) produces only an $n$-th mode response ($x >
  0$). The iteration procedure will add a coupled part to the solution of each
  mode. 

  \nomenclature[a-E]{$E^\ast$}{Complex error function}
  \nomenclature[c-PDE]{PDE}{Partial Differential Equation}
  \nomenclature[a-Phi0]{$\Phi_{n^{\prime}}^{(i)}$}{Elements of vector $\boldsymbol{\Phi}^{(i)}$}
  \nomenclature[g-mun]{$\mu_{n^{\prime}}$}{Transformed mixed wavenumber of mode $n^{\prime}$}
  \nomenclature[a-Kn]{$K_{n^{\prime}}$}{Transformed acoustic wavenumber of mode $n^{\prime}$}
  \nomenclature[a-Phi]{$\bar{\Phi}_{n^\prime}$}{Transformed second-part scattered modal potential}

  As discussed above, by substituting $\boldsymbol{\Phi}^{(0)}$ into the coupling terms on the right hand side of (\ref{equ:MatrixFormWaveEquations}), one obtains
  \begin{equation}
    \mathcal{D}\boldsymbol{\Phi} = \boldsymbol{A} \boldsymbol{\Phi} + \boldsymbol{B}\frac{\partial \boldsymbol{\Phi}^{(0)}}{\partial x}.
    \label{iteration1st}
  \end{equation}
  However, it should be noted that due to the inhomogeneous nature of these equations, they cannot be solved using the standard Schwarzschild technique. One therefore needs to manipulate these equations so that they change to homogeneous ones. Note that $\boldsymbol{\Phi}^{(0)}$ satisfies (\ref{initial}), hence, for $x \ne 0$, where $\boldsymbol{\Phi}^{(0)}$ is first-order continuously differentiable, the following equation holds:
  \begin{equation}
    \mathcal{D}\frac{\partial\boldsymbol{\Phi}^{(0)}}{\partial x} = \boldsymbol{A} \frac{\partial\boldsymbol{\Phi}^{(0)}}{\partial x}.
    \label{initial_derivative}
  \end{equation}

  Making use of (\ref{initial_derivative}), one can show that (\ref{iteration1st}) can be equivalently written as
  \begin{equation}
    \mathcal{D} (\boldsymbol{\Phi} + \bm{\alpha}\frac{\partial \boldsymbol{\Phi}^{(0)}}{\partial x}) = \boldsymbol{A}(\boldsymbol{\Phi} + \bm{\alpha} \frac{\partial \boldsymbol{\Phi}^{(0)}}{\partial x}),
    \label{equ:Homogenous1st}
  \end{equation}
  where $\bm{\alpha}$ is a coefficient matrix whose entries are
  \begin{equation}
    \alpha_{ml} = \frac{B_{ml}}{k_{2m}^2 - k_{2l}^2} =
    \Bigg\{
      \begin{aligned}
        &\frac{-4h}{\pi^2(m-l)^2}, &m-l = odd\\
        &0, &m-l = even.
      \end{aligned}
      \label{}
    \end{equation}

It should be noted that for $z = 0$ (\ref{equ:Homogenous1st}) only holds when
$x \ne 0$, not for $x \in \boldsymbol{R}$. In order to apply the
Schwarzschild's technique, it must be valid over the whole domain. However,
since the singularity of $\partial \boldsymbol{\Phi}^{(0)}/\partial x$ only
exists at $x = 0$, similar to the differentiation of $H(y)$, we may again make
use of the generalized function to account for this singularity. Let $\partial
\boldsymbol{\hat{\Phi}}^{(0)}/\partial x$ denote the generalized
differentiation, which allows the presence of generalized functions at singular
point $ x= 0$ but equals to $\partial \boldsymbol{\Phi}^{(0)}/\partial x$
elsewhere, then equation
    \begin{equation}
      \mathcal{D}\frac{\partial\boldsymbol{\hat{\Phi}}^{(0)}}{\partial x} = \boldsymbol{A} \frac{\partial\boldsymbol{\hat{\Phi}}^{(0)}}{\partial x}
      \label{equ:ModifiedP0}
    \end{equation}
    needs to hold for any $x \in \boldsymbol{R}$. The Schwarzschild technique suggests that if (\ref{equ:ModifiedP0}) does hold, then the routine application of the steps described in (\ref{equ:initialEle}) to (\ref{equ:initialResults}) shall recover the value of $\partial \boldsymbol{\Phi}^{(0)}/\partial x$ for $x > 0$. Thus, one can show that the intended $\partial \boldsymbol{\hat{\Phi}}^{(0)}/\partial x$ can indeed be found as
    \begin{equation}
      \frac{\partial \hat{\Phi}_{n^{\prime}}^{(0)}}{\partial x} (x,0)=
      \frac{\partial \Phi_{n^{\prime}}^{(0)}}{\partial x}(x,0) - \Phi_{ia}
      a_{n^\prime}(1+\i)(-\sqrt{\mu_{n^{\prime}}}) \sqrt{-2\pi x} \delta(x),
      \label{Fexpression}
    \end{equation}
    where $\partial \hat{\Phi}_{n^{\prime}}^{(0)}/\partial x$ denotes the element of $\partial \boldsymbol{\hat{\Phi}}^{(0)}/\partial x$ corresponding to the $n^{\prime}$-th mode and
    \begin{equation}
	\int_{-\infty}^0 \delta(x) \ud x = \frac{1}{2}.
      \label{equ:DeltaDefinition}
    \end{equation}
    Now, the first iterated solution, $\boldsymbol{\Phi}^{(1)}$, can be obtained by solving the following equation
    \begin{equation}
	\mathcal{D} (\boldsymbol{\Phi} +
	\bm{\alpha} \frac{\partial \boldsymbol{\hat{\Phi}}^{(0)}}{\partial x} ) =
	    \boldsymbol{A}(\boldsymbol{\Phi} +
	    \bm{\alpha} \frac{\partial \boldsymbol{\hat{\Phi}}^{(0)}}{\partial x}),
      \label{equ:IterationEqu}
    \end{equation}
    using the steps described in (\ref{equ:initialEle}) to (\ref{equ:initialResults}).
Continuing this iteration process gives $\boldsymbol{\Phi}^{(2)}$, $\boldsymbol{\Phi}^{(3)}$, \ldots, and the exact solutions $\boldsymbol{\Phi}_t$ after adding the initial potential field can be expressed as
    \begin{equation}
      \boldsymbol{\Phi}_t(x,0) = \boldsymbol{N}(x) + \boldsymbol{C}^{(1)}(x)+ \boldsymbol{C}^{(2)}(x)+ \boldsymbol{C}^{(3)}(x)+ \cdots,
      \label{equ:Solutions}
    \end{equation}
where the non-coupled part is denoted by $\boldsymbol{N}$, while the coupled parts are denoted by $\boldsymbol{C}^{(i)}$ ($i = 1, 2, 3\cdots$). Here only the entries of $\boldsymbol{N}$ and $\boldsymbol{C}^{(1)}$ corresponding to mode $n^{\prime}$ are presented, which are
    \begin{equation}
      N_{n^{\prime}}(x) = - \Phi_{ia} \e^{\i k_1x} a_{n^{\prime}} (1+\i)E^\ast(\mu_{n^{\prime}}x)
      \label{equ:NonscatteringPart}
    \end{equation}
    and
    \begin{equation}
      \begin{aligned}
        C_{n^{\prime}}^{(1)}(x) = - \Phi_{ia}\e^{\i k_1x}(1+\i)
	\sum_{m=-\infty}^\infty\alpha_{n^{\prime}m}a_m \Bigg(& \i k_1 \Big( E^\ast(\mu_{n^{\prime}}x)-E^\ast(\mu_m x)\Big)\\
        &\negmedspace{} +\sqrt{\frac{\mu_m}{2\pi
	x}}\Big(\e^{-\i\mu_{n^{\prime}}x} - \e^{-\i\mu_m x} \Big)\Bigg)
      \end{aligned}
      \label{equ:ScatteringPart1}
    \end{equation}
    respectively. The result of the second iteration is rather complex and is provided in Appendix A. It is worth noting that due to the
    iterative nature of this solution, the function $\boldsymbol{C}^{(i)}$ becomes more and more complex as $i$ increases. However, if $\boldsymbol{C}^{(i)}$
    vanishes sufficiently quickly, the higher orders can be dropped without causing significant errors. This appears to be the case for the frequency range of interest considered in this paper.

    \nomenclature[a-Phihat]{$\boldsymbol{\hat{\Phi}}^{(0)}$}{Generalized potential in vector form}
    \nomenclature[g-alpha]{$\bm{\alpha}$}{Coefficient matrix in the first-order results}
    \nomenclature[g-xi]{$\xi$}{Integral variable}

    \nomenclature[a-N]{$\boldsymbol{N}$}{Non-coupled part of solution in $\boldsymbol{\Phi}_t$}
    \nomenclature[a-C]{$\boldsymbol{C}^{(i)}$}{Coupled part of solution in $\boldsymbol{\Phi}_t$}
    \nomenclature[a-N]{$N_{n^{\prime}}$}{Elements of vector $\boldsymbol{N}$}
    \nomenclature[a-C]{$C_{n^{\prime}}^{(i)}$}{Elements of vector $\boldsymbol{C}^{(i)}$}

    Substituting (\ref{equ:NonscatteringPart}) and (\ref{equ:ScatteringPart1}) into (\ref{equ:Solutions}), a first-order approximation of the exact solution is obtained.
    The induced potential due to the gust interaction is finally obtained by summing the modal solutions over all different modes and transforming  back to the original physical coordinate system, namely
    \begin{equation}
      \begin{aligned}
        \Phi_t(x^{\prime},y^{\prime},0) =\sum_{n^{\prime}=-\infty}^{\infty}
	[N_{n^{\prime}}+C^{(1)}_{n^{\prime}} + C^{(2)}_{n^{\prime}}+ \cdots]
	(x^{\prime}-H(y^{\prime}),0) \e^{\i k_{2n^{\prime}}y^{\prime}},
        \label{equ:ScatteredSurfacePressure1st}
      \end{aligned}
    \end{equation}
    where the $N_{n^{\prime}}$ and $C^{(1)}_{n^{\prime}}$ functions are defined in (\ref{equ:NonscatteringPart}) and (\ref{equ:ScatteringPart1}) respectively, $C^{(2)}_{n^{\prime}}$ can be found in the appendix and the terms in the second parenthesis are the arguments for the $N_{n^\prime}$ and $C^{(i)}_{n^\prime} \,(i = 1, 2, 3, \cdots)$ functions.

    As shown in (\ref{equ:ScatteredSurfacePressure1st}), the induced potential
    field can now be expressed in terms of an infinite series. In a limiting
    case, when $h = 0$, all the $C^{(i)}_{n^{\prime}}\,(i = 1, 2, 3\cdots)$
    terms on the right hand side of (\ref{equ:ScatteredSurfacePressure1st})
    vanish, and Amiet's formulation is recovered. An interesting fact about the
    solution developed here is the the proportionality of $C^{(i)}_{n^\prime} \propto h^i$ ($i = 1, 2, 3, \cdots$), and thus (\ref{equ:ScatteredSurfacePressure1st})
    may be understood as a perturbation solution with respect to $h$. It can be
    shown easily that at sufficiently low frequencies, i.e. $k_1 h <
    \pi^2 /4$, the infinite series is convergent. At higher frequencies, the
    series still appears to be convergent, but it can be expected that for a
    proper approximation, a higher truncation number and higher-order
    iterations will be required. The convergence issue will be further
    discussed in more details in Section~\ref{sec:comparison}. 

 \subsubsection{Induced far-field sound pressure}
 
 As shown in the previous section, the induced potential field in the time domain can be found as
    \begin{equation}
      \begin{aligned}
        \phi_t(x^{\prime},y^{\prime},0, t) =\sum_{n^{\prime}=-\infty}^{\infty}
	[N_{n^{\prime}}+C^{(1)}_{n^{\prime}} + C^{(2)}_{n^{\prime}}+ \cdots]
	(x^{\prime}-H(y^{\prime}),0) \e^{\i k_{2n^{\prime}}y^{\prime}} \e^{-\i\omega t}.
        \label{equ:InducedPotentialTime}
      \end{aligned}
    \end{equation}
    The pressure field is related to the velocity potential through the momentum equation, as
    \begin{equation}
      p = -\rho_0 U (\frac{\partial \phi_t}{\partial x^\prime} - \i k_1\phi_t),
      \label{}
    \end{equation}
    therefore, the pressure jump $\Delta p$  across the flat plate is given by
    \begin{equation}
      \begin{aligned}
        \Delta p(x^{\prime},y^{\prime},0, t) =
	2\sum_{n^{\prime}=-\infty}^{\infty}
	[P^{(0)}_{n^{\prime}}+P^{(1)}_{n^{\prime}} + P^{(2)}_{n^{\prime}}+
	\cdots] (x^{\prime}-H(y^{\prime}),0) \e^{\i k_{2n^{\prime}}y^{\prime}}
	\e^{-\i\omega t},
      \end{aligned}
      \label{equ:PressureJump}
    \end{equation}
    where
    \begin{equation}
      P^{(0)}_{n^{\prime}}(x) =  \rho_0 U \Phi_{ia} (1+\i)\e^{\i k_1x}
      a_{n^{\prime}} \sqrt{\mu_{n^\prime}} \frac{1}{\sqrt{2\pi x}}\e^{-\i\mu_{n^\prime}x}
      \label{equ:P0}
    \end{equation}
    and
    \begin{IEEEeqnarray*}{rCll}
      P_{n^{\prime}}^{(1)}(x) & = & \IEEEeqnarraymulticol{2}{l}{\rho_0 U
	  \Phi_{ia}(1+\i) \e^{\i k_1x}} \\
      && \sum_{m=-\infty}^\infty& \alpha_{n^{\prime}m}a_m \Bigg[\i k_1
	  \frac{1}{\sqrt{2\pi x}} \Big( \sqrt{\mu_{n^\prime}}
	  \e^{-\i\mu_{n^\prime}x}- \sqrt{\mu_m} \e^{-\i\mu_m x}\Big)\\
      &&& \negmedspace{} - i \sqrt{\frac{\mu_m}{2\pi x}}\Big(\mu_{n^\prime}
  \e^{-\i\mu_{n^{\prime}}x} - \mu_m \e^{-\i\mu_m x} \Big) - \frac{1}{2}
  \sqrt{\frac{\mu_m}{2\pi x}} \frac{1}{x} \Big(\e^{-\i\mu_{n^{\prime}}x} -
  \e^{-\i\mu_m x} \Big)\Bigg]. \IEEEeqnarraynumspace \IEEEyesnumber
      \label{equ:P1}
    \end{IEEEeqnarray*}
The second order solution, $P_{n^{\prime}}^{(2)}(x)$, is given in the
appendix. Having obtained the pressure jump across the flat plate, the
far-field sound can be found using the surface pressure
integral~\citep{Amiet1975}, as
    \begin{equation}
      p_f(\boldsymbol{x}, \omega) = \frac{-\i\omega x_3}{4\pi c_0 S_0^2} \iint_s
      \Delta P(x^{\prime}, y^{\prime}) \e^{-\i kR} \ud x^{\prime} \ud y^{\prime},
      \label{equ:SerratedFarField}
    \end{equation}
    where $S_0^2 = x_1^2 + \beta^2(x_2^2 + x_3^2)$ denotes the stretched
    distance due to the mean flow, the pressure jump has a harmonic form
    $\Delta P \e^{-\i\omega t}= \Delta p$ and the radiation distance $R$ takes the form of
    \begin{equation}
      R= \frac{M_0(x_1 - x^{\prime})-S_0}{\beta^2} + \frac{x_1x^{\prime} + x_2 y^{\prime}\beta^2}{\beta^2S_0}.
      \label{equ:PhaseRelation}
    \end{equation}

    \nomenclature[a-S]{$S_0$}{$S_0 = \sqrt{x_1^2 + \beta^2(x_2^2 + x_3^2)}$}
    \nomenclature[a-R]{$R$}{Function denoting phase relation in (\ref{equ:PhaseRelation})}
    \nomenclature[a-p]{$\Delta p$}{Pressure jump across the flat plate}

    By substituting the solution obtained in (\ref{equ:PressureJump}) into (\ref{equ:SerratedFarField}), the far-field sound pressure can be found as
    \begin{equation}
      p_f(\boldsymbol{x},\omega,k_2) = 2 \rho_0 U \Phi_{ia}
      \mathcal{L}(\omega,k_1, k_2) \left(\frac{-\i\omega x_3}{4\pi c_0 S_0^2}\right)\lambda\frac{\sin\big((N+1/2)\lambda(k_2-kx_2/S_0)\big)}{\sin\big(\lambda/2(k_2-kx_2/S_0)\big)}.
      \label{equ:pressure}
    \end{equation}
    Here, $2N+1$ represents the number of sawteeth along the span and the non-dimensional far-field sound gust-response function $\mathcal{L}$ is defined as
    \begin{equation}
	\begin{aligned}
      \mathcal{L}(\omega, k_1, k_2) = & (1+\i)\frac{1}{\lambda} \left(\sum_{n^{\prime
      }=-\infty}^\infty \big(\Theta_{n^{\prime}}^{(0)} +
      \Theta_{n^{\prime}}^{(1)} +\Theta_{n^{\prime}}^{(2)}+\cdots
      \big)\right) \times\\
     & \e^{-\i k/\beta^2(M_0x_1-S_0)} \e^{\i k/\beta^2(M_0-x_1/S_0)h},
	\end{aligned}
      \label{equ:SerratedResponseFunction}
    \end{equation}
    with
    \begin{equation}
      \begin{aligned}
        &\Theta_{n^{\prime}}^{(0)} =  a_{n^{\prime}} \sqrt{\mu_{n^\prime}} S_{n^{\prime}n^{\prime}},\\
        & \Theta_{n^{\prime}}^{(1)} =
	\sum_{m=-\infty}^\infty\alpha_{n^{\prime}m}a_m \Bigg[\i k_1  \Big( \sqrt{\mu_{n^\prime}} S_{n^\prime n^\prime} - \sqrt{\mu_m} S_{n^\prime m}\Big) \\
	& \quad \quad \quad \quad - \i \sqrt{\mu_m}\Big(\mu_{n^\prime} S_{n^\prime n^\prime} - \mu_m S_{n^\prime m} \Big) -  \sqrt{\mu_m}  \Big( T_{n^\prime n^\prime} -  T_{n^\prime m} \Big)\Bigg],
      \end{aligned}
    \end{equation}
    and $\Theta_{n^{\prime}}^{(2)}$ terms are provided in the Appendix. The function $S_{nm}$ and $T_{nm}$ in the above equations are given by
    \begin{equation}
      \begin{aligned}
        S_{nm} = \sum_{j=0}^1& \frac{1}{\i\kappa_{nj}}\Bigg\{
	    \frac{1}{\sqrt{\eta_{Am}}}\Big[\e^{\i\kappa_{nj}
	    \lambda_{j+1}}E^\ast(\eta_{Am}(c-\epsilon_{j+1})) - \e^{\i\kappa_{nj} \lambda_{j}}E^\ast(\eta_{Am}(c-\epsilon_{j}))\Big]\\
        &  - \frac{1}{\sqrt{\eta_{Bnmj}}} \e^{\i\kappa_{nj} (\lambda_j + (c-\epsilon_j)/\sigma_j)}\Big[E^\ast(\eta_{Bnmj}(c-\epsilon_{j+1})) - E^\ast(\eta_{Bnmj}(c-\epsilon_{j}))\Big] \Bigg\},
      \end{aligned}
      \label{equ:S_nm}
    \end{equation}
    \begin{equation}
      \begin{aligned}
        T_{nm} = \sum_{j=0}^1 & \frac{1}{\i\kappa_{nj}}  \Bigg\{
	    \frac{-\i\eta_{Am}}{\sqrt{\eta_{Am}}} \Big[\e^{\i\kappa_{nj}
	    \lambda_{j+1}}E^\ast(\eta_{Am}(c-\epsilon_{j+1})) - \e^{\i\kappa_{nj} \lambda_{j}}E^\ast(\eta_{Am}(c-\epsilon_{j}))\Big]\\
        &\ \;   + \frac{\i\eta_{Bnmj}}{\sqrt{\eta_{Bnmj}}} \e^{\i\kappa_{nj}
    (\lambda_j + (c-\epsilon_j)/\sigma_j)}\Big[
    E^\ast(\eta_{Bnmj}(c-\epsilon_{j+1})) -
E^\ast(\eta_{Bnmj}(c-\epsilon_{j}))\Big] \Bigg\}, 
      \end{aligned}
      \label{equ:T_nm}
    \end{equation}
    where
    \begin{equation}
      \begin{aligned}
        &\kappa_{nj} = k_{2n} - kx_2/S_0 + k/\beta^2(M_0-x_1/S_0)\sigma_j,\\
        &\eta_{Am} = - K_m + kM_0/(\beta^2 + \sigma^2) - k/\beta^2(M_0-x_1/S_0),\\
        &\eta_{Bnmj}= - K_m + kM_0/(\beta^2 + \sigma^2) + (k_{2n} - kx_2/S_0)/\sigma_j.
      \end{aligned}
    \end{equation}
  
\subsection{General-gust solution}
\label{subsec:general-gust}
Equation (\ref{equ:pressure}) gives the far-field sound pressure induced by a
single gust of form $w_i = w_{ia}\e^{-\i(\omega t-k_1 x^{\prime}-k_2 y^{\prime})}$. For a more general incoming gust given by (\ref{equ:generalGust}), the induced far-field pressure can be obtained from
\begin{equation}
    \begin{aligned}
  p_f(\boldsymbol{x},\omega) = &2 \rho_0 \left(\frac{-\i\omega x_3}{4\pi c_0
  S_0^2}\right) \\
  &\int_{-\infty}^{\infty}\frac{-\tilde{w}(\omega /U, k_2)}{\gamma_d(\omega/U, k_2))} \mathcal{L}(\omega,\omega/U, k_2) \lambda\frac{\sin\big((N+1/2)\lambda(k_2-kx_2/S_0)\big)}{\sin\big(\lambda/2(k_2-kx_2/S_0)\big)} \ud k_2.
  \label{equ:generalPressure}
    \end{aligned}
\end{equation}
where $p_f$ is the far-field sound pressure and $\gamma_d(k_1, k_2) = \sqrt{(k_1 \beta + k M_0/\beta)^2 + k_2^2 - (k / \beta)^2}$. The far-field sound PSD, $S_{pp}(\boldsymbol{x}, \omega)$, can then be found from   
\begin{IEEEeqnarray}{rCl}
    S_{pp}(\boldsymbol{x},\omega) &=& \lim_{T \to \infty} \frac{\pi}{T}\overline{p_f(\boldsymbol{x},\omega) p_f^\ast(\boldsymbol{x},\omega)}\nonumber \\ 
  &=& \left(\frac{\rho_0 \omega x_3}{2 \pi c_0 S_0^2}\right)^2 U \nonumber\\
  &\int_{-\infty}^{\infty}&\frac{\Phi_{ww}(\omega/U, k_2)}{|\gamma_d(\omega/U,
  k_2)|^2} |\mathcal{L}(\omega,\omega/U, k_2)|^2 \lambda^2
  \frac{\sin^2\big((N+1/2)\lambda(k_2-kx_2/S_0)\big)}{\sin^2\big(\lambda/2(k_2-kx_2/S_0)\big)}
  \ud k_2, \IEEEeqnarraynumspace
  \label{equ:PSD}
\end{IEEEeqnarray}
where the overbar and star denote the ensemble average and complex conjugate
respectively and we have made use of the fact that~\citep{Amiet1975}
\begin{equation}
  \overline{w(\omega/U, k_2) w^\ast(\omega/U, k_2^\prime)} = \frac{L}{\pi} \delta(k_2 - k_2^\prime) \Phi_{ww}(\omega/U, k_2).
\end{equation}
where $\Phi_{ww}$ denotes the wavenumber power spectral density of the incoming vertical
fluctuation velocity. Equation~(\ref{equ:PSD}) can be further simplified by noting that when the span $d$ of the flat plate is large, we have
\begin{equation}
  \lambda^2 \frac{\sin^2\big((N+1/2)\lambda(k_2-kx_2/S_0)\big)}{\sin^2\big(\lambda/2(k_2-kx_2/S_0)\big)} \sim  2 \pi d \sum_{m = -\infty}^{\infty} \delta(k_2 - k x_2/S_0 + 2 m \pi / \lambda),
\end{equation}
and (\ref{equ:PSD}) becomes 
\begin{IEEEeqnarray}{rCl}
    S_{pp}(\boldsymbol{x}, \omega) &=& \lim_{T \to \infty} \frac{\pi}{T}\overline{p_f(\boldsymbol{x},\omega) p_f^\ast(\boldsymbol{x},\omega)}\nonumber \\ 
  &=& \left(\frac{\rho_0 \omega x_3}{2 \pi c_0 S_0^2}\right)^2 U (2 \pi d) \nonumber\\ 
  &&\sum_{m = -\infty}^{\infty} \frac{\Phi_{ww}(\omega/U, k x_2 / S_0 + 2 m \pi
  /\lambda)}{|\gamma_d(\omega/U, k x_2 / S_0 + 2 m \pi / \lambda)|^2}
  |\mathcal{L}(\omega,\omega/U, k x_2 / S_0 + 2 m \pi /\lambda)|^2.
  \IEEEeqnarraynumspace
  \label{equ:SimpliedPSD}
\end{IEEEeqnarray}
Equation~(\ref{equ:SimpliedPSD}) is the fundamental equation of this paper. For the cases with the observer located on the mid-span plane, i.e. $x_2 = 0$, the sound pressure PSD reduces to
\begin{IEEEeqnarray}{rCl}
    S_{pp}(\boldsymbol{x}, \omega) = (2 \pi d U)\left(\frac{\rho_0 \omega
    x_3}{2\pi c_0 S_0^2}\right)^2 \sum_{m = -\infty}^{\infty} \frac{\Phi_{ww}(\omega/U, 2 m \pi /\lambda)}{|\gamma_d(\omega/U, 2 m \pi / \lambda)|^2} |\mathcal{L}(\omega,\omega/U, 2 m \pi /\lambda)|^2.
  \IEEEeqnarraynumspace
  \label{equ:MidSpanSimpliedPSD}
\end{IEEEeqnarray}
It is worthy of pointing out that (\ref{equ:MidSpanSimpliedPSD}) shows that the
far-field sound PSD has a linear dependence on the incoming turbulence
intensities. The quantity $\Phi_{ww}$ can be obtained from various models for
turbulence energy spectrum, such as but not limited to Von Karman spectrum
model. Moreover, it should be emphasized that due to the assumption of a
uniform mean flow and a flat plate, the effects of lifting potential flow
around a realistic aerofoil are neglected in this paper. It has been shown in
several papers that when the angle of attack and camber are not zero, these
effects are important~\citep{Tsai1990,Myers1995,Myers1997}. However, these
effects would diminish as the Mach number decreases~\citep{Myers1995}.
Consequently, the developed model in this paper should be a good approximation
when at low angle of attacks and/or the Mach number is low. More importantly,
though these lifting-flow effects can change the overall far-field sound
spectra, it is unlikely to cause any significant changes to the predicted sound
reduction, which is perhaps more important and of more practical interests.

\section{Comparison with experiments}
\label{sec:comparison} Having obtained an analytical solution for the far-field
sound PSD, (\ref{equ:MidSpanSimpliedPSD}), we can now compare the results
against the experimental data and also carry out a parametric study and
investigate the effect of the serration geometry and turbulence parameters on
the noise generation mechanism. As seen in (\ref{equ:MidSpanSimpliedPSD}), the
model requires a prior knowledge of the wavenumber power spectral density of
incoming vertical fluctuation velocity ($\Phi_{ww}$). Previous experiments on
leading edge noise have shown that the turbulent upwash velocity spectra can
well be captured by the Von Karman energy spectrum
model~\citep{Amiet1975,Narayanan2015}. By adopting the Von Karman energy
spectrum model, \citet{Amiet1975} showed that
\begin{equation}
  \Phi_{ww}(k_1, k_2) = \frac{4 \bar{u^2}}{9 \pi k_e^2} \frac{\hat{k}_1^2 + \hat{k}_2^2}{(1 + \hat{k}_1^2 + \hat{k}_2^2)^{7/3}},
  \label{equ:Phiww}
\end{equation}
where $u$ denotes the streamwise fluctuating velocity and $k_e$, $\hat{k}_1$ and $\hat{k}_2$ are given by
\begin{equation}
    k_e = \frac{\sqrt{\pi} \Gamma(5/6)}{L_t \Gamma(1/3)}, \hat{k}_1 =
    \frac{k_1}{k_e}, \hat{k}_2 = \frac{k_2}{k_e},
\end{equation}
where $L_t$ is the integral scale of the turbulence and $\Gamma$ is the Gamma function. 

\begin{figure}
  \begin{subfigure}{0.495\textwidth}
  \centering
  \includegraphics[width=\linewidth]{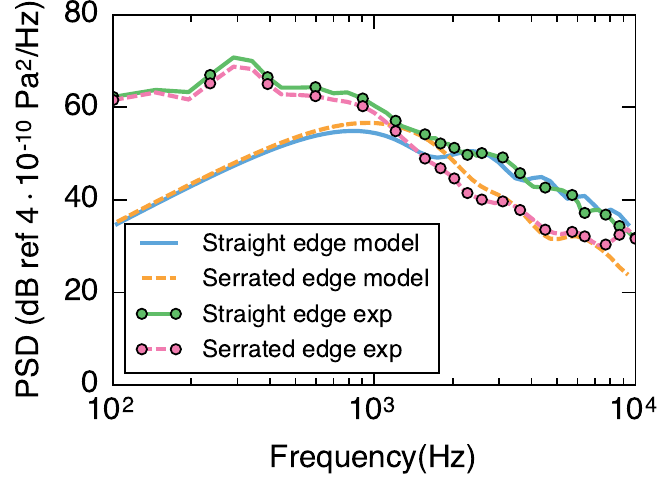}
  \caption{$h / c = 0.067$.}
  \label{fig:validation1}
\end{subfigure}
\begin{subfigure}{0.495\textwidth}
  \centering
  \includegraphics[width=\linewidth]{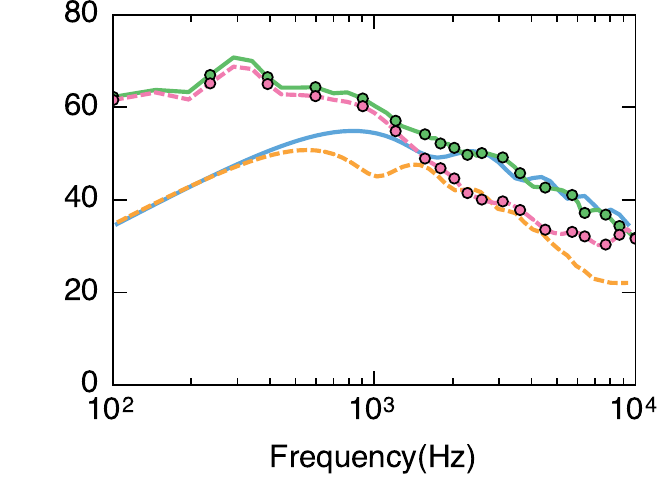}
  \caption{$h / c = 0.167$.}
  \label{fig:validation2}
\end{subfigure}
\caption{The validation of the second-order model, (\ref{equ:MidSpanSimpliedPSD}), with experimental data for the baseline (blue) and serrated (red) flat plates.
    The observer is at $90^\circ$ above the flat plate in the mid-span plane.
    The Von Karman model for isotropic turbulence is used with the mean flow
    velocity of $U = 60 \text{ m/s}$, integral length-scale of $L_t = 0.006
    \text{m}$ and turbulence intensity of $2.5\%$, as measured
    by~\citet{Narayanan2015}. Both serrations have a spanwise wavelength of $\lambda / c = 0.067$.}
\label{fig:validation}
\end{figure}

In order to validate the new model, we compare the far-field noise predictions,
using (\ref{equ:MidSpanSimpliedPSD}), against the experimental data measured
by~\citet{Narayanan2015}. The experiment was carried out for a flat plate
immersed in a turbulent flow with the mean flow velocity of $U = 60\text{
m/s}$, turbulent intensity of about $2.5\%$ and streamwise integral scale of
$L_t = 0.006 \text{\ m}$. The flat plate had a mean-chord length of $c \approx
0.175 \text{ m}$ and span length of $d = 0.45 \text{ m}$, fitted with a
sinusoidal serration with the spanwise wavelength of $\lambda / c = 0.067$ and
amplitudes of $h / c = 0.067$ (figure~\ref{fig:validation1}) and $h / c =
0.167$ (figure~\ref{fig:validation2}). Note that though in our model a sawtooth
serration is used, for such a sharp serration, we expect the differences
between the two serration profiles to be negligible. Note also that, in the
experiment, the microphones were positioned outside the jet flow in the
far-field. However, as pointed by \cite{Amiet1975}, the shear of the jet mean
flow has no refraction effects for the observer directly above the flat plate,
i.e. $90^\circ$ above the flat plate leading-edge. The convection effects of
the ambient mean flow, as considered in the model, have an order of $\beta^2$
for such an observer. Since the Mach number in the experiment was very low
(less than $0.2$), the convection effects of mean flow can be safely neglected.
Therefore, we can proceed to compare the sound spectra measured in the
experiment to the results obtained in the model. As mentioned earlier, the Von
Karman velocity spectrum, (\ref{equ:Phiww}), was used to represent the
wavenumber power spectral density of the vertical fluctuation velocity. The
high level of noise at low frequencies observed in the experimental data, as
mentioned by~\citet{Narayanan2015}, is believed to be due to the dominance of
the open-jet wind tunnel background noise and also the grid-generated vortex
shedding and its interaction with the flat plate. Therefore, the disagreement
at low frequencies between the experiment and model prediction, as shown in
figure~\ref{fig:validation1}, is believed to be due to the dominance of jet
noise. Another possible reason for such a disagreement is the perfect-coherence
assumption in the streamwise direction, which will be described in details in
subsequent sections. In the mid to high frequencies, however, the model
provided excellent agreement with measured data. In particular, there exists a
perfect match between the predicted and observed peaks of the leading edge
noise of the serrated flat plate at frequencies above $f>1000 \text{ Hz}$.
This suggests that the model captures the essential physics and gives accurate
prediction of the noise from plates and aerofoil fitted with leading-edge
serrations.  Predictions under the same flow condition for a much sharper
serrations, i.e.  $\lambda / c = 0.067$ and $h / c = 0.167$, are presented in
figure~\ref{fig:validation2}. It can be seen that sharper serrations are more
effective in reducing leading edge noise, which has been demonstrated by both
the experimental data and model predictions. The agreement between the model
and experiments continues to be very good in the frequency range where leading
edge noise is dominant. The slight mismatch at high frequencies ($f > 7 \text{
kHz}$) for the serrated case is likely to be caused by other noise mechanisms
present in the experiment, as seen in the experimental data, such as the
trailing-edge noise.

\nomenclature[a-Lt]{$L_t$}{Integral scale of the turbulence}
\nomenclature[a-L]{$L$}{Large number used for spacial Fourier transformation}
\begin{figure}
    \centering
    \includegraphics[width=0.6\textwidth]{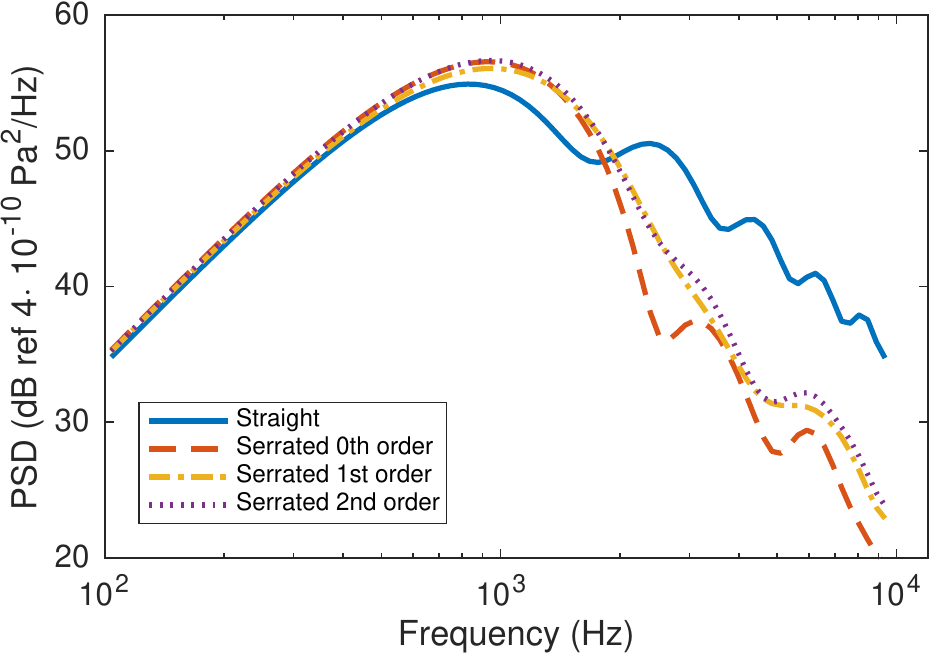}
    \caption{The convergence of the 0th, 1st and 2nd order solutions.}
    \label{fig:L}
\end{figure}

The results presented in figure \ref{fig:validation} were calculated using the second-order solutions, but we have not yet examined the rate of convergence of the solution in (\ref{equ:MidSpanSimpliedPSD}). To demonstrate that the second-order solution can provide a sufficiently accurate solution, we present the predicted sound pressure spectra for the first validation case (figure~\ref{fig:validation1}) using the 0th, 1st and 2nd-order solutions. The results are shown in figure~\ref{fig:L}. Though a difference of up to 5 dB can be observed between the 0th- and 1st-order solutions, the difference between the 1st- and 2nd-order solutions is uniformly less than 1 dB over the entire frequency range of interest. This suggests that the second-order solution should serve as a good approximation for the serration cases considered in this study and over the frequency range of interest.  
 
\section{Discussions}
\subsection{Effects of serration geometry}
\label{subsec:effects}
In this section, we carry out a parametric numerical evaluation of the model to study the effects of serration geometry and Mach number on leading edge noise. Since we are primarily interested in the effects of serration geometry and flow convective effects, we shall use the same chord-length and incoming turbulence statistical quantities as in the preceding section, i.e. $c \approx 0.175\text{ m}$, $L_t = 0.006\text{ m}$ and turbulent intensity of $2.5\%$. Results will be presented for the far-field sound power spectra at $90^\circ$ above the flat plate in the mid-span plane. 

\begin{figure}
    \centering
    \begin{subfigure}{0.495\textwidth}
	\centering
	\includegraphics[width=\linewidth, trim=0 15 0 0]{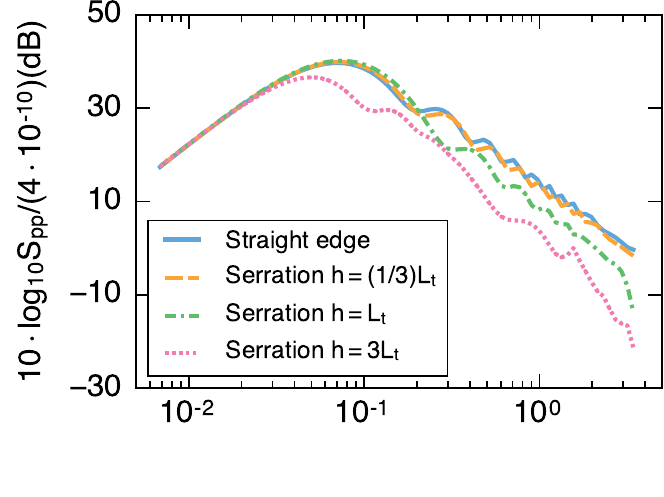}
	\caption{$\lambda = 3 L_t,\, M_0 = 0.1$}
	\label{fig:parametricWideM1}
    \end{subfigure}
    \begin{subfigure}{0.495\textwidth}
	\centering
	\includegraphics[width=\linewidth, trim=0 15 0 0]{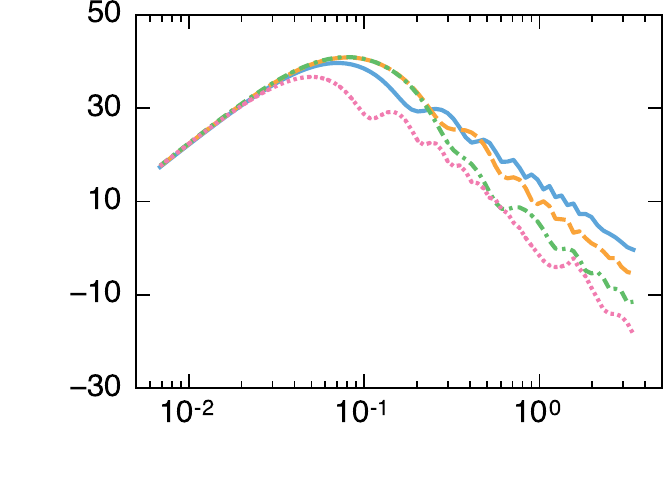}
	\caption{$\lambda =  L_t,\, M_0 = 0.1$}
	\label{fig:parametricMidM1}
    \end{subfigure}
    \begin{subfigure}{0.495\textwidth}
	\centering
	\includegraphics[width=\linewidth, trim=0 15 0 0]{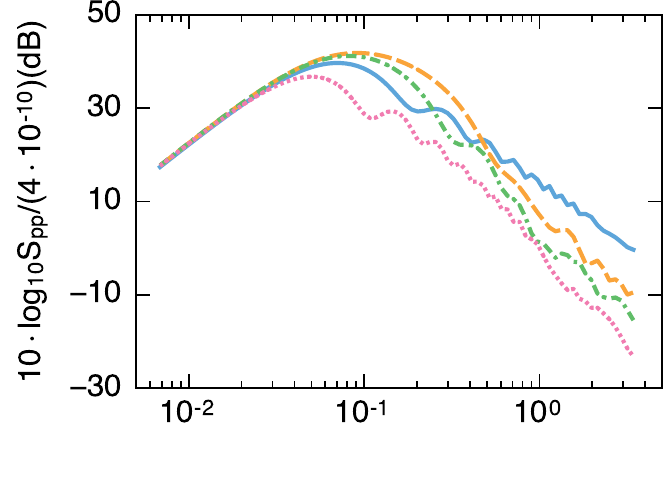}
	\caption{$\lambda = (1 / 3) L_t,\, M_0 = 0.1$}
	\label{fig:parametricSharpM1}
    \end{subfigure}
    \begin{subfigure}{0.495\textwidth}
	\centering
	\includegraphics[width=\linewidth, trim=0 15 0 0]{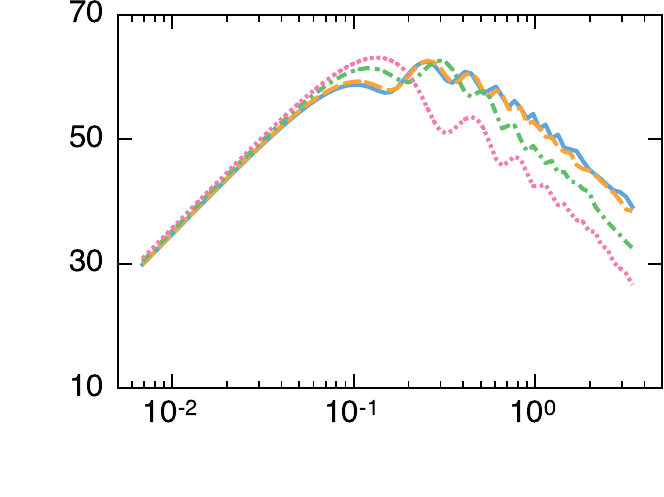}
	\caption{$\lambda = 3 L_t,\, M_0 = 0.4$}
	\label{fig:parametricWideM4}
    \end{subfigure}
    \begin{subfigure}{0.495\textwidth}
	\centering
	\includegraphics[width=\linewidth]{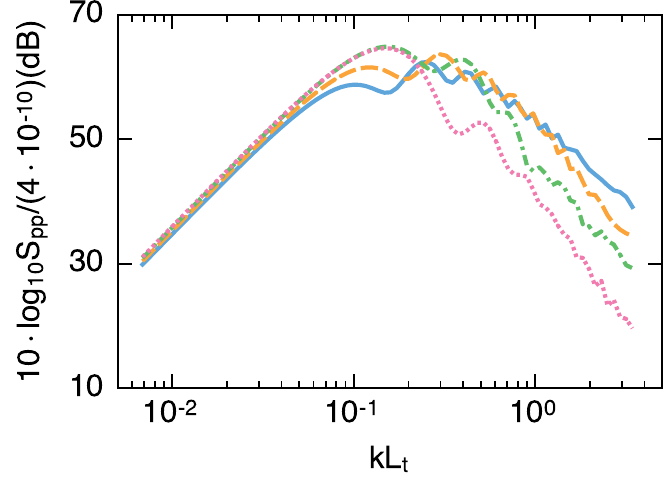}
	\caption{$\lambda =  L_t,\, M_0 = 0.4$}
	\label{fig:parametricMidM4}
    \end{subfigure}
    \begin{subfigure}{0.495\textwidth}
	\centering
	\includegraphics[width=\linewidth]{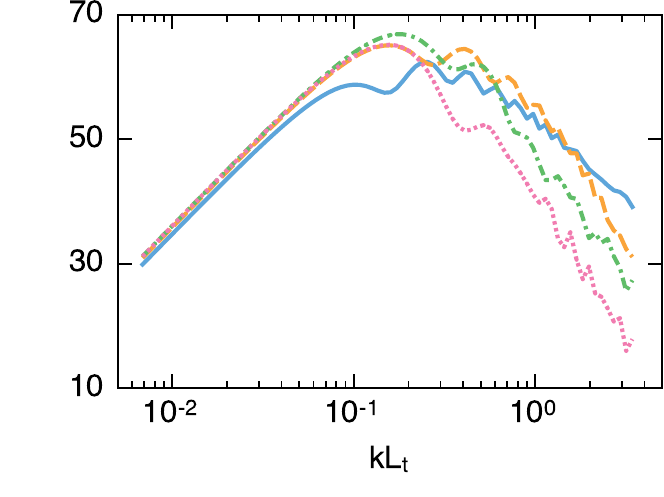}
	\caption{$\lambda = (1 / 3) L_t,\, M_0 = 0.4$}
	\label{fig:parametricSharpM4}
    \end{subfigure}
    \caption{The effects of varying $h$ and $\lambda$ on far-field sound
	spectrum. The observer is located at $90^\circ$ above the flat plate in
	the mid-span plane. The Von Karman model for isotropic turbulence is
	used with the integral length-scale of $L_t = 0.006 \text{m}$ and
	turbulence intensity of $2.5\%$, as measured by~\citet{Narayanan2015}. The flat plate has a mean chord length of $c = 0.175 \text{ m}$ and span length of $d = 0.45\text{ m}$.}
    \label{fig:parametricStudy}
\end{figure}

Results in figure~\ref{fig:parametricWideM1} presents the far-field sound
pressure level for a flat plate with wide serrations, i.e. $\lambda = 3 L_t$,
with different amplitudes, i.e. $h$ from $1/3 L_t$ to $3 L_t$, in a relatively
low Mach number flow of $M_0 = 0.1$. Note here, we use ``wide'' to describe the
serrations which have longer wavelength $\lambda$ compared to those shown
subsequently. Similarly, ``sharp'' serrations have been used to describe
serrations with small value of $\lambda$. It is clear from the results that
serrations of $h = 1/3 L_t$ have virtually no effects on reducing the far-field
sound. This is expected since a leading edge with wide and short serrations
will act very similar to a straight edge. On the other hand, serrations with
relatively large root-to-tip amplitude, $h = L_t$, start to reduce the noise at
high frequencies, as shown by the dash-dotted line in
figure~\ref{fig:parametricWideM1}. The use of longer serrations, i.e. $h = 3
L_t$, is shown to result in significant reduction of the far-field noise, even
at low frequencies. 

Figure~\ref{fig:parametricMidM1} shows the results for a flat plate with
leading-edge serration wavelength of $\lambda = L_t$. The effect of varying the
serration amplitude ($h$) on the reduction of turbulence interaction noise is
similar to those observed before. However, an important difference compared
with the wider serration case with $\lambda = 3 L_t$, see
figure~\ref{fig:parametricWideM1}, is that an area of noise increase appears at
low frequencies for short serrations (small $h$). The reason of such noise
increases at low frequencies will be discussed later in Section
\ref{sec:NoiseReductionMech}. Results are also presented for very sharp
serrations ($\lambda = 1/3 L_t$), see figure~\ref{fig:parametricSharpM1}. The
noise increase observed at low frequencies for serrations with small $h$ is now
more pronounced. However, one can see that so long as the serration is long
enough, the noise increase disappears completely and that using sharp
serrations (small $\lambda$) results in more effective sound reduction at high
frequencies compared with the wide serrations. In summary, the results suggest
that in order to suppress the leading edge noise, the serration wavelength
$\lambda$ has to be sufficiently small and the root-to-tip amplitude has to be
large.

The effect of flow convective effects, particularly at high Mach numbers, can
also be studied using the new model. As mentioned earlier, since the
formulations are based on Amiet's leading-edge noise theory, where the
convection effects of the uniform mean flow has properly accounted for, the new
model should suffer no constraints caused by high-speed mean flow convection
effects. Therefore the model can be used for higher Mach numbers (when the
uniform mean flow assumption is permissible). This capability is particularly
important as most of the experimental data available are collected at low Mach
numbers of up to $0.23$~\citep{Narayanan2015}.
Figures~\ref{fig:parametricWideM4} to \ref{fig:parametricSharpM4} present the
far-field noise from a flat plate in a turbulent flow with the Mach number of
$M_0 = 0.4$. Results are presented for wide ($\lambda = 3 L_t$) to sharp
($\lambda =1/3 L_t$) serrations. In general, the results show the same trends
as before, that is greater noise reduction can be achieved using sharp
serrations and that the use of wide and small leading edge serrations can lead
to noise increase at about $k L_t=0.1$. The level of noise increase for shallow
serrations (large $\lambda$ and small $h$) has been observed to increase
significantly with Mach number. 

\subsection{Directivity} 
The effects of leading edge serrations on far-field noise directivity have also
been investigated. In this section, we shall study the effects of a specific
leading edge serration on the directivity patterns at different
non-dimensionlized frequencies ($kL_t$) and Mach numbers
\begin{figure}
    \centering
    \begin{subfigure}{0.4\textwidth}
	\centering
	\includegraphics[width=\linewidth]{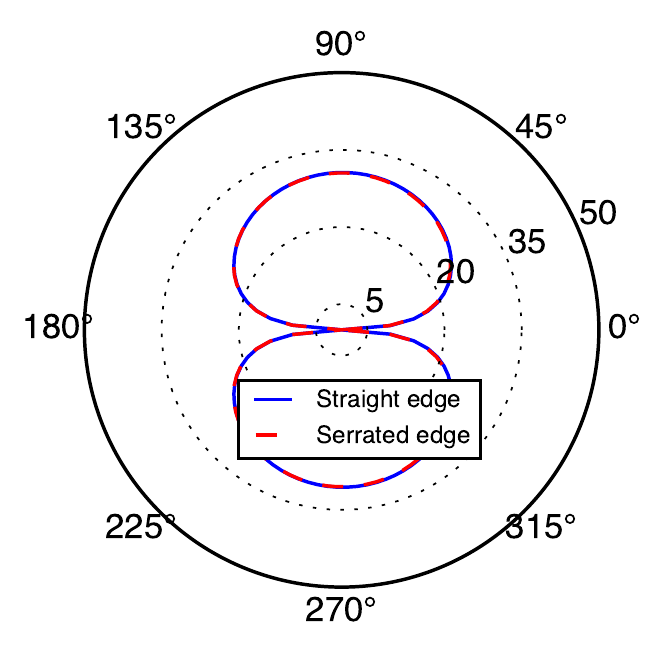}
	\caption{$kL_t = 0.02$}
	\label{fig:directivityFreq1M1}
    \end{subfigure}
    \begin{subfigure}{0.4\textwidth}
	\centering
  \includegraphics[width=\linewidth]{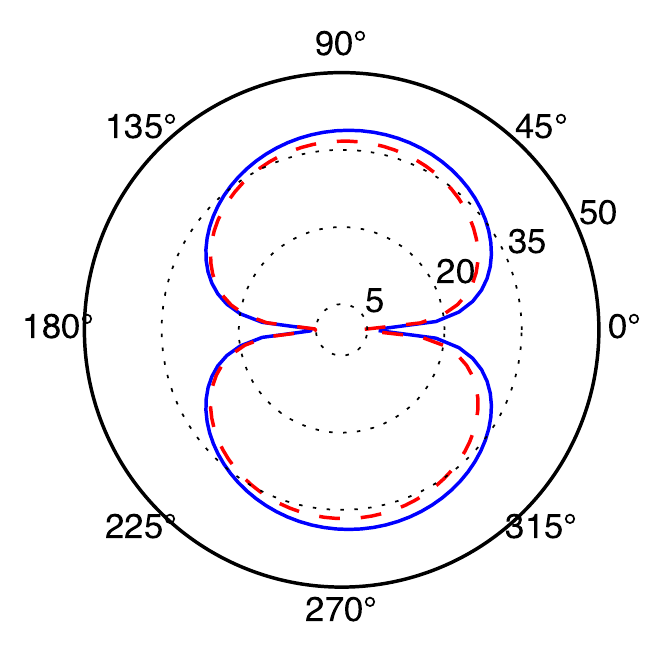}
  \caption{$kL_t = 0.05$}
  \label{fig:directivityFreq2M1}
\end{subfigure}
  \begin{subfigure}{0.4\textwidth}
  \centering
  \includegraphics[width=\linewidth]{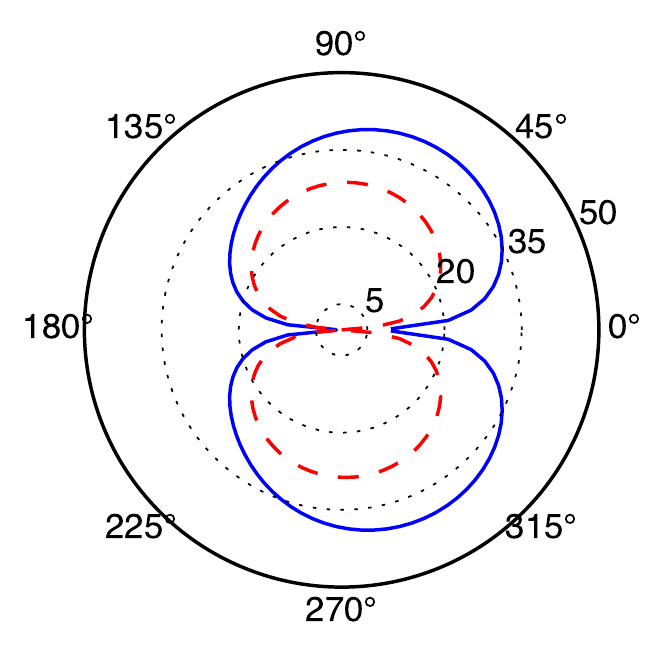}
  \caption{$kL_t = 0.1$}
  \label{fig:directivityFreq3M1}
\end{subfigure}
\begin{subfigure}{0.4\textwidth}
  \centering
  \includegraphics[width=\linewidth]{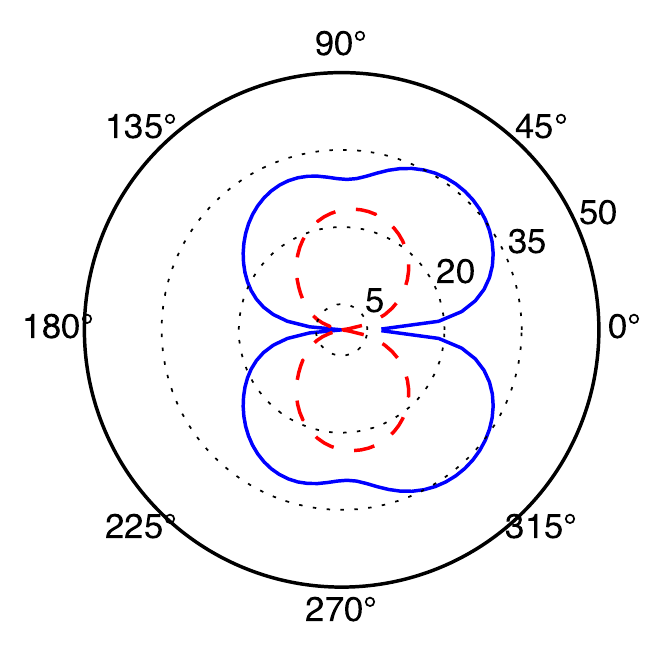}
  \caption{$kL_t = 0.2$}
  \label{fig:directivityFreq4M1}
\end{subfigure}
  \begin{subfigure}{0.4\textwidth}
  \centering
  \includegraphics[width=\linewidth]{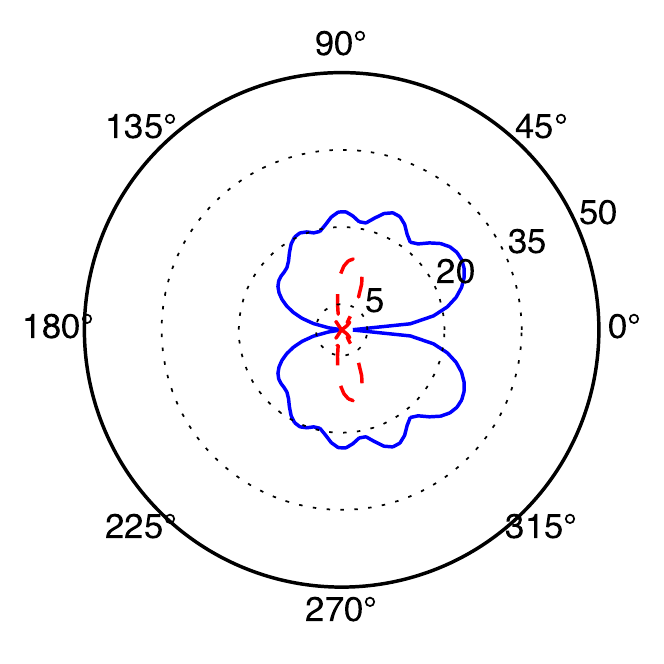}
  \caption{$kL_t = 0.5$}
  \label{fig:directivityFreq5M1}
\end{subfigure}
\begin{subfigure}{0.4\textwidth}
  \centering
  \includegraphics[width=\linewidth]{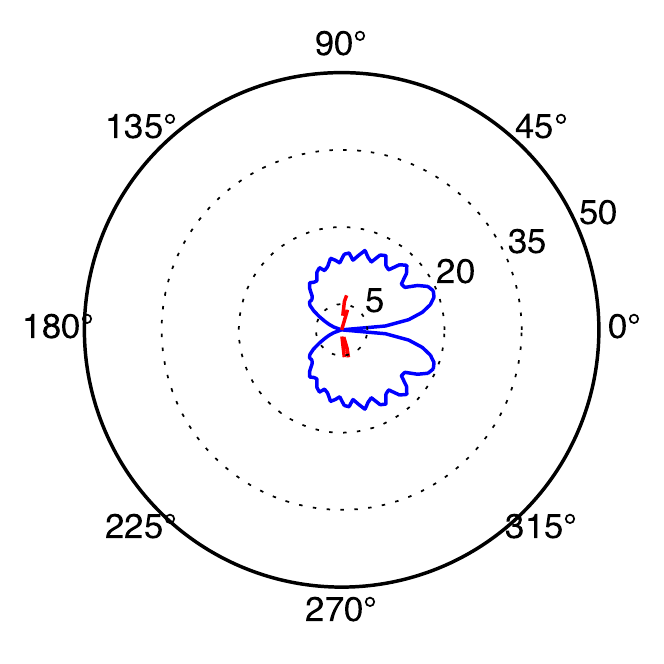}
  \caption{$kL_t = 1$}
  \label{fig:directivityFreq6M1}
\end{subfigure}
\caption{The directivity of leading edge noise for both straight and serrated
    edges at Mach number $M_0 = 0.1$ in the mid-span plane. The Von 
    Karman model for isotropic turbulence is used in the analytical model with the 
    turbulent intensity of $2.5\%$ and length-scale of $L_t = 0.006 \text{ m}$.
    Results are presented at different frequencies ($kL_t$), 
    corresponding to convective wavenumbers of 
    $k_1 L_t = 0.2, 0.5, 1, 2, 5$ and $10$.} 
\label{fig:DirectivityM1}
\end{figure}
\begin{figure} \centering \begin{subfigure}{0.4\textwidth} \centering
	\includegraphics[width=\linewidth]{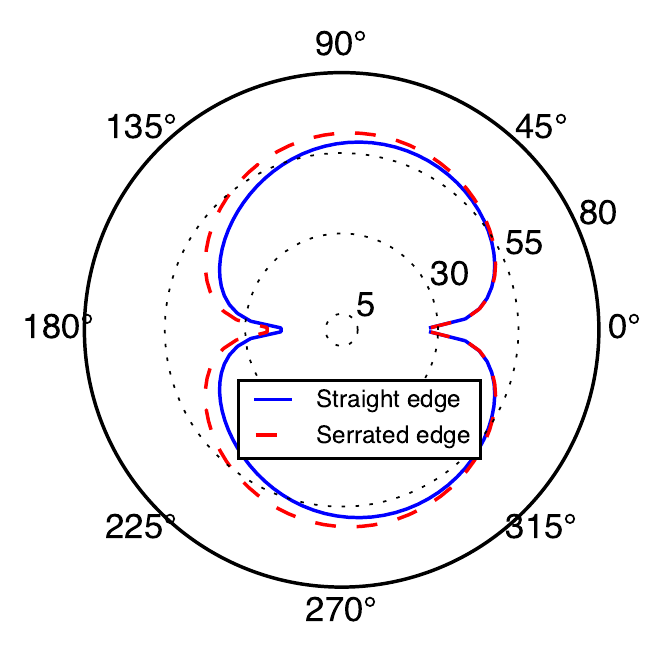}
	\caption{$kL_t = 0.08$} \label{fig:directivityFreq1M4} \end{subfigure}
    \begin{subfigure}{0.4\textwidth} \centering
	\includegraphics[width=\linewidth]{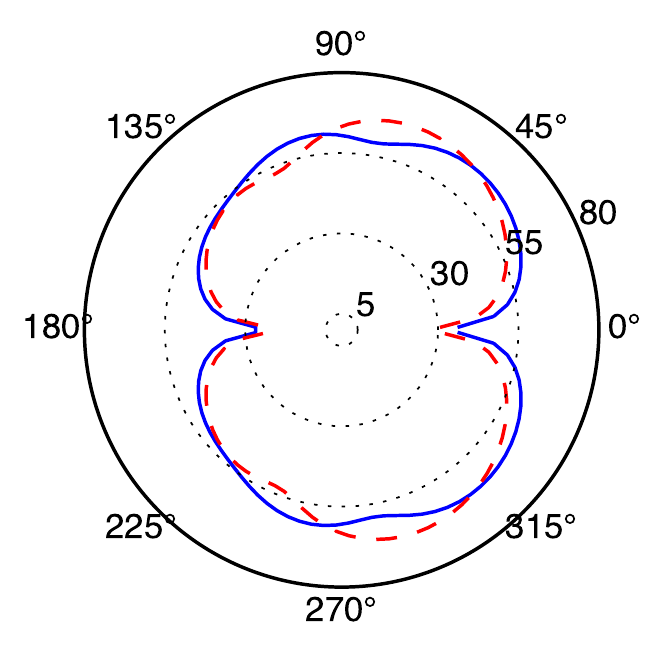}
	\caption{$kL_t = 0.2$} \label{fig:directivityFreq2M4} \end{subfigure}
    \begin{subfigure}{0.4\textwidth} \centering
	\includegraphics[width=\linewidth]{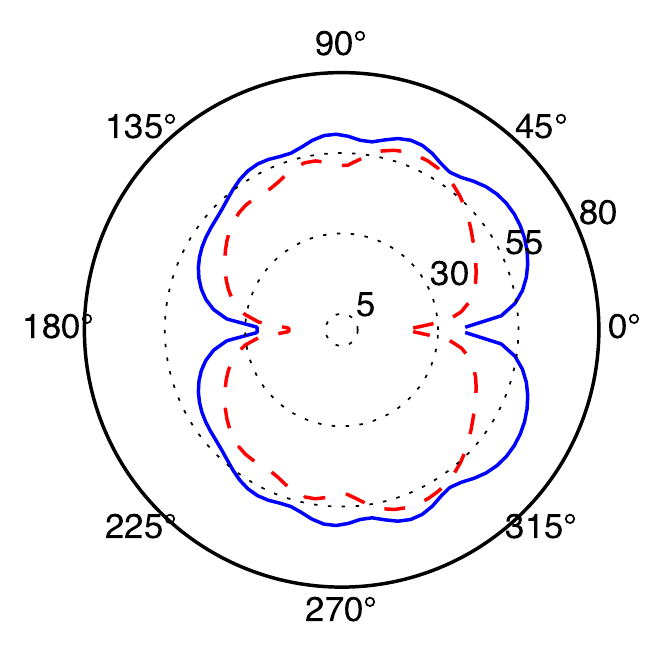}
	\caption{$kL_t = 0.4$} \label{fig:directivityFreq3M4} \end{subfigure}
    \begin{subfigure}{0.4\textwidth} \centering
	\includegraphics[width=\linewidth]{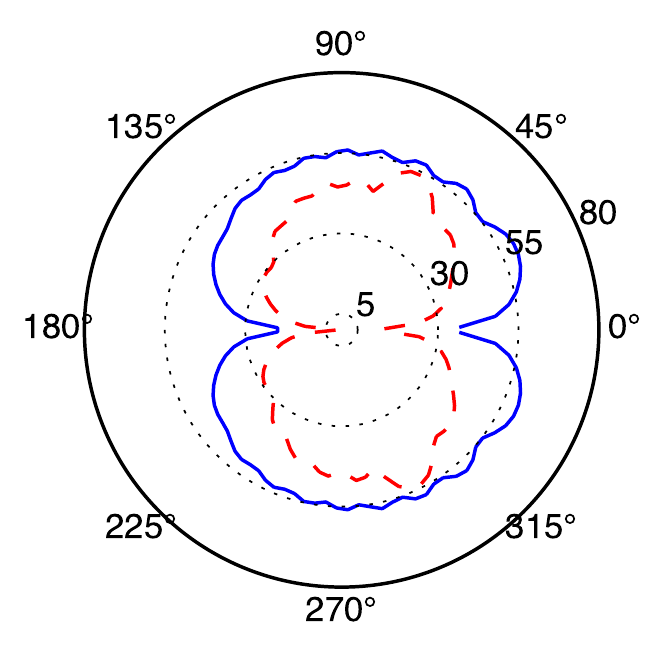}
	\caption{$kL_t = 0.8$} \label{fig:directivityFreq4M4} \end{subfigure}
    \begin{subfigure}{0.4\textwidth} \centering
	\includegraphics[width=\linewidth]{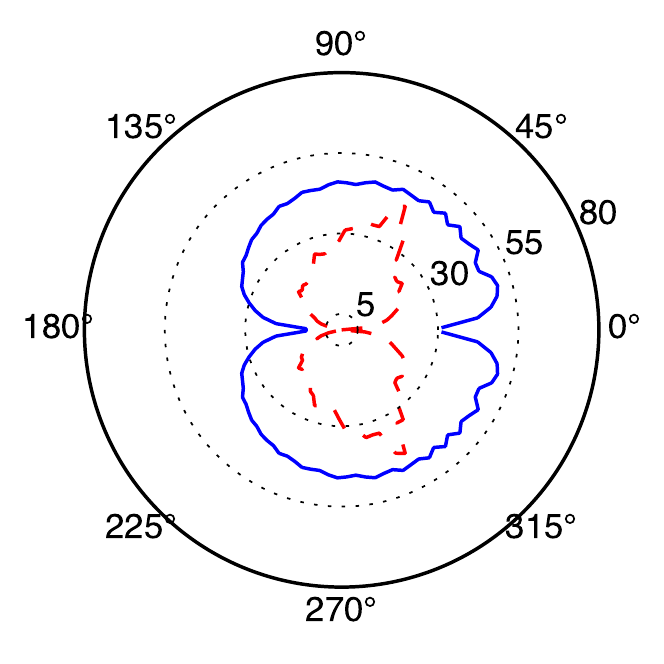}
	\caption{$kL_t = 2$} \label{fig:directivityFreq5M4} \end{subfigure}
    \begin{subfigure}{0.4\textwidth} \centering
	\includegraphics[width=\linewidth]{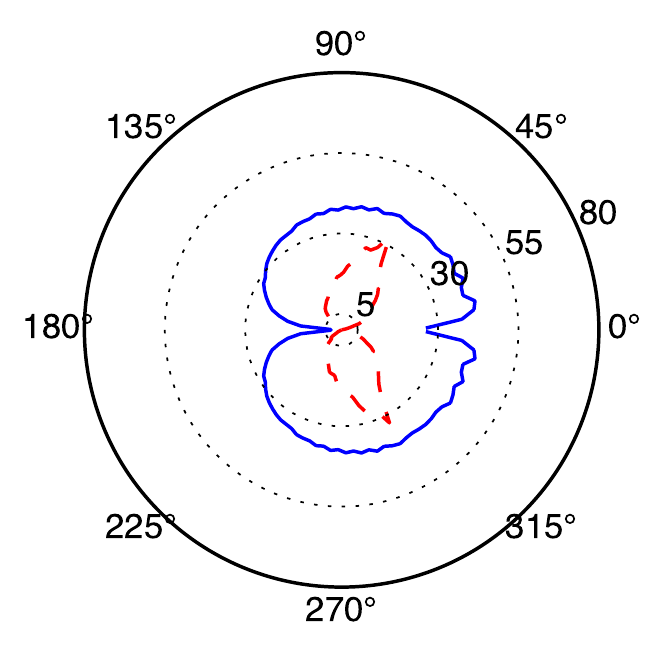}
	\caption{$kL_t = 4$} \label{fig:directivityFreq6M4} \end{subfigure}
    \caption{The directivity of leading edge noise for both straight and
	serrated edges at Mach number $M_0 = 0.4$ in the mid-span plane. 
	The Von Karman model for isotropic turbulence is used in the
	analytical model with the turbulent intensity of $2.5\%$ and
	length-scale of $L_t = 0.006 \text{ m}$. Results are presented at
    different frequencies ($kL_t$), corresponding to convective wavenumbers of
$k_1 L_t = 0.2, 0.5, 1, 2, 5$ and $10$.} \label{fig:DirectivityM4}
\end{figure}
($M_0 = 0.1$ and $0.4$).  From Section~\ref{subsec:effects}, we see that in
order to achieve significant noise reduction, the serration wavelength
$\lambda$ has to be sufficiently small and the serration amplitude $2h$ has to
be sufficiently large.  However, it is also found that when $\lambda$ is too
small, e.g.  $\lambda = (1/3) L_t$, there is also a large noise increase at
intermediate frequencies. Therefore, in practical applications, the serration
profile with $\lambda = L_t$ and $h = 3L_t$ is preferred. In this section, we
choose this geometry to study the effects of serration on directivity at
different frequencies and Mach numbers.  As before, the flat plate has a
mean-chord length of $c = 0.175 \text{ m}$ and span length $d = 0.45\text{ m}$.
Also, the incoming flow turbulence intensity is set to $2.5\%$ and integral
length-scale is taken as $L_t = 0.006 \text{ m}$. Results are presented at
different acoustic wavenumbers $kL_t$, corresponding to the convective
wavenumbers of $k_1 L_t = 0.2, 0.5, 1, 2, 5$ and $10$ (see figures
\ref{fig:DirectivityM1}).


Figure~\ref{fig:directivityFreq1M1} shows the directivity patterns for both the
straight and serrated edges at the frequency of $kL_t = 0.02$ ($k_1 L_t =
0.2$). At such low frequencies, the serrations have no effects on the radiated
sound. Slight noise reduction only appears when frequency increases to $kL_t =
0.05$ ($k_1 L_t = 0.5$), as shown in figure~\ref{fig:directivityFreq2M1}.
Further increasing the frequency results in more effective sound reduction, as
shown in figure~\ref{fig:directivityFreq3M1}. It is interesting to note that at
mid-frequencies, i.e. $kL_t = 0.2$ and $0.5$ ($k_1 L_t = 2$ and $5$),
greater noise reduction is obtained for observer locations closer to the
trailing edge side of the flat plate, i.e. $\theta=0^\circ$, see
figures~\ref{fig:directivityFreq4M1} and \ref{fig:directivityFreq5M1}. At
higher frequency $kL_t  = 1$ ($k_1 L_t =10$), significant noise reduction can
be achieved at all radiation angles and the noise reduction at
$\theta=90^\circ$ reaches up to $10\text{ dB}$. Another interesting phenomenon
observed is the significant changes to the directivity pattern of the radiated
noise. As seen for the low frequencies, i.e. $kL_t = 0.02$ and $0.05$,
the introduction of leading edge serrations does not particularly change the
dipolar shape of the radiated sound field. However, at higher frequencies, the
cardioid shape of leading edge noise is changed to a more tilted dipolar shape,
as seen in figures~\ref{fig:directivityFreq3M1} to \ref{fig:directivityFreq6M1}.

\begin{figure}
    \centering
    \begin{subfigure}{0.4\textwidth}
	\centering
	\includegraphics[width=\linewidth]{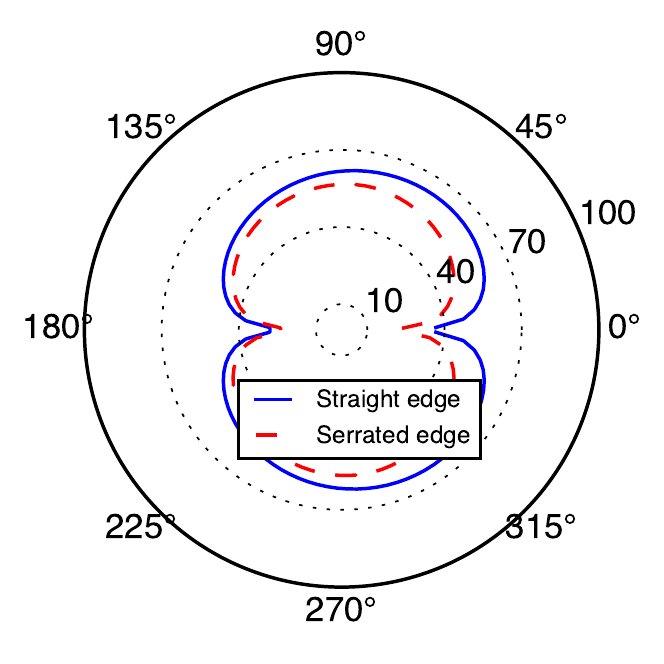}
	\caption{$M_0 = 0.1$}
    \end{subfigure}
    \begin{subfigure}{0.4\textwidth}
	\centering
	\includegraphics[width=\linewidth]{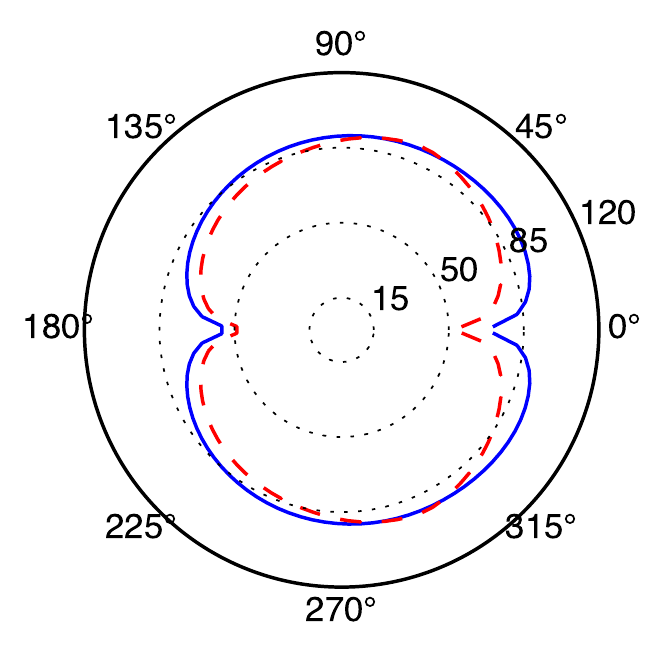}
	\caption{$M_0 = 0.4$}
    \end{subfigure}
    \caption{The leading edge far-field noise OASPL directivity patterns for both straight and
    serrated leading edges in the mid-span plane. The serration used has a wavelength of $\lambda = L_t$ and half root-to-tip amplitude of $h = 3L_t$ and
the incoming flow has a turbulent intensity of $2.5\%$ and integral length-scale of $L_t = 0.006 \text{m}$}
    \label{fig:OASPLdirectivity}
\end{figure}

The far-field noise directivity results at Mach number of $M_0 = 0.4$ are
presented in figure~\ref{fig:DirectivityM4}.
Figure~\ref{fig:directivityFreq1M4} shows the directivity results at $kL_t =
0.08$ ($k_1 L_t = 0.2$). Compared to figure~\ref{fig:directivityFreq1M1}
for $M_0 = 0.1$, we see a slight noise increase at low frequencies. This is
consistent with our findings in the preceding section, i.e. noise
increase is more pronounced at high Mach numbers. This noise increase persists
at around $90^\circ$ above and below the plate for frequencies up to $kL_t =
0.2$ ($k_1 L_t = 0.5$), as shown in figure~\ref{fig:directivityFreq2M4}.
The use of leading edge serrations at $kL_t = 0.4$ ($k_1 L_t = 1$),
however, leads to noise reduction at all angles, as can be seen in
figure~\ref{fig:directivityFreq3M4}. Figures~\ref{fig:directivityFreq4M4} to
\ref{fig:directivityFreq6M4} show the directivity patterns at $kL_t = 0.8$, $2$
and $4$, respectively, corresponding to convective wavenumber of $k_1 L_t
= 2, 5,$ and $10$. As before, results here confirm that the leading edge
serrations are more effective at high frequencies and that the use of leading
edge serration changes the far-field noise directivity from a cardioid shape to
a tilted dipole.

In addition to the sound pressure level at different frequencies, the overall
sound pressure level (OASPL) results can also provide some insight into the
total sound energy radiated at different angles and the effects of leading edge
serrations. This also provides an opportunity to better understand the
effectiveness of the leading edge serration as most of the OASPL results
obtained experimentally are contaminated due to the strong low frequency
background jet noise contribution. The OASPL results here are obtained by
integrating the sound power of the frequency range of $kL_t = 0.02$ to $2$ (for
$M_0 = 0.1$) and $kL_t = 0.04$ to $4$ (for $M_0 = 0.4$). As in the previous
section, the serration used has a wavelength of $\lambda = L_t$ and half
root-to-tip amplitude of $h = 3L_t$ and the incoming flow has a turbulent
intensity of $2.5\%$ and integral length-scale of $L_t = 0.006 \text{m}$.
Results are presented at Mach numbers of $M_0 = 0.1$ and $0.4$, see
figures~\ref{fig:DirectivityM1} and \ref{fig:DirectivityM4}. Results have shown
that the use of leading edge serration can result in significant reduction of
the OASPL of up to 5-10 dB. It has also been observed that the leading edge
serration are more effective at low Mach numbers and small polar angles,
i.e.downstream towards to the plate's surface. This was also observed in the
far-field SPL results in figures \ref{fig:DirectivityM1} and
\ref{fig:DirectivityM4} at mid to high frequencies.

\subsection{Noise reduction mechanism} \label{sec:NoiseReductionMech}

Inspection of the equations developed in Section~\ref{sec:Formulation} shows
that in order for the leading edge serration treatment to be effective, two
geometrical criteria must be met. The detailed derivations are not provided for
the sake brevity, and interested readers can refer to the paper by the
authors~\citep{Lyu2016} on noise from aerofoils with trailing edge serrations.
The two geometrical criteria are (1) $ \omega h / U\gg 1$ and (2) $\omega h_e
/U \gg 1$, where the effective half root-to-tip length $h_e$ is defined by $h_e
= \sigma l_{y^\prime} / 2$ and $l_{y^\prime}$ is the spanwise correlation
length defined by 
\begin{equation*}
  l_{y^\prime} (\omega) = \frac{1}{R(\omega, 0)}\int_{-\infty}^{\infty}
    R(\omega, y^\prime) \ud y^\prime,
\end{equation*}
where $R(\omega, y^\prime)$ is the spanwise two-point correlation of the
incoming turbulent velocity. Using Von Karman model, \citet{Amiet1975}
showed that $l_{y^\prime}(\omega)$ can be obtained as 
\begin{equation}
    l_{y^\prime}(\omega) = \frac{16L_t}{3}
    \left(\frac{\Gamma(1/3)}{\Gamma(5/6)}\right)^2
    \frac{(\omega/Uk_e)^2}{\left(3 + 8 (\omega/Uk_e)^2\right) \sqrt{1 + (\omega/Uk_e)^2}}.
    \label{}
\end{equation}
It may seem somewhat unexpected to obtain the
dependence of the sound reduction on the spanwise correlation length.  A
careful inspection of (\ref{equ:PSD}) reveals that this dependence originates
from the wavenumber power spectral density $\Phi_{ww}(\omega, k_2)$ in
(\ref{equ:PSD}). The inverse Fourier transformation of
$\Phi_{ww}(\omega, k_2)$ over $k_2$ yields the two-point cross spectrum
$R(\omega, y^\prime)$.  The first criterion is consistent with the findings in
the numerical work by~\citet{Lau2013} and some experimental data available for
different leading edge serrations~\citep{Chaitanya2016}. The second criterion,
although less discussed in recent works, is an important condition and relates
the serration geometry to the structure of the incoming turbulence.

As shown in other numerical and experimental works, the main cause of noise
reduction is due to the destructive sound interferences caused by the serration. In
that sense, the noise reduction mechanism is very similar to that of trailing
edge serration, as previously discussed by~\citet{Lyu2016}.
The better understanding of the destructive sound interference phenomenon may help us better explain the physical implications of the two above-mentioned criteria. The first
condition $\omega h /U \gg 1$ is to ensure that a complete phase variation (of minimum $2\pi$) of the scattered
pressure along the serration edge is achieved. At low frequencies, this
requires the serration amplitude $2h$ to be large in order to achieve significant sound reduction, as only a large value of $h$ can ensure a complete phase variation of the
scattered pressure field along the serration edge. For the cases when the
serration amplitude $h$ is not large enough, there would be little
variation of the scattered pressure, i.e. in-phase radiation along the edge. This can result in noise increase at low frequencies, as observed in figures~\ref{fig:validation}
and~\ref{fig:parametricStudy}, which is due to the fact that the wetted edge by an in-phase pressure field for the serrated edge is longer than that of the straight-edge.
It should be noted that in this paper we assume that the incoming turbulence is frozen. This would imply a perfect spatial correlation in the streamwise direction for the upwash velocity field. For serrations with large $h$, the assumption of perfect streamwise correlation of the turbulence becomes inappropriate, and therefore the condition developed may not hold. This might explain why a noise increase is predicted in figure~\ref{fig:validation1} while
this did not happen in experimental results.
However, for mid- to
\begin{figure}
  \centering
  \includegraphics[]{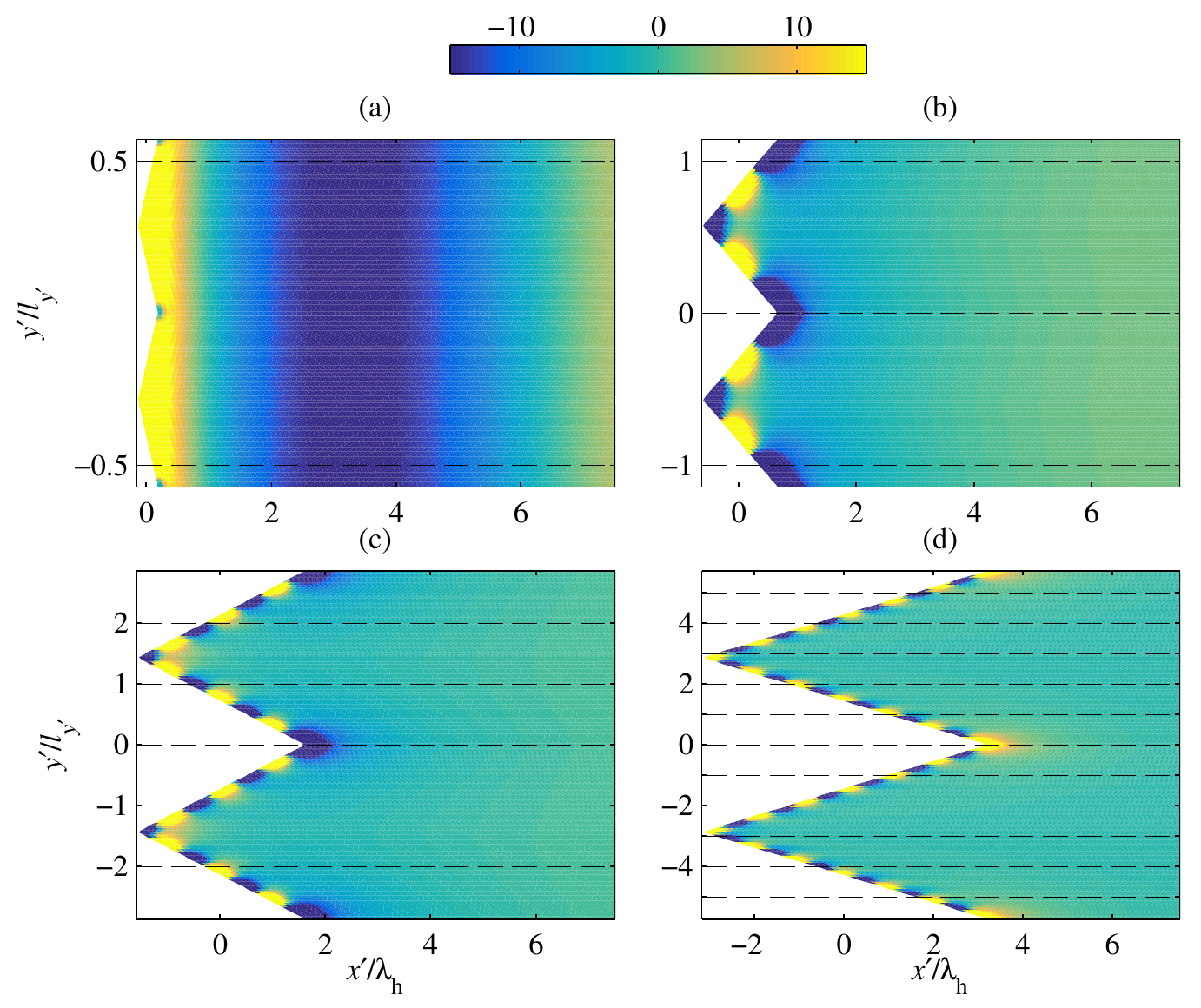}
  \caption{The effects of varying $\omega h / U$ on the scattered surface
  pressure at a fixed frequency and $\omega h_e /U = 7$: a) $\omega h / U= 1$;
  b) $\omega h /U  = 4$; c) $\omega h / U = 10$; d) $\omega h / U = 20$. The
  color shows the normalized scattered pressure on the upper surface of the
  flat plate. The
  vertical axis shows the spanwise coordinate normalized by the spanwise
  correlation length and the horizontal axis shows the streamwise coordinate normalized by the hydrodynamic wavelength.}
  \label{fig:surfacePressureSerratedFreq}
\end{figure}
\begin{figure}
  \centering
  \includegraphics[]{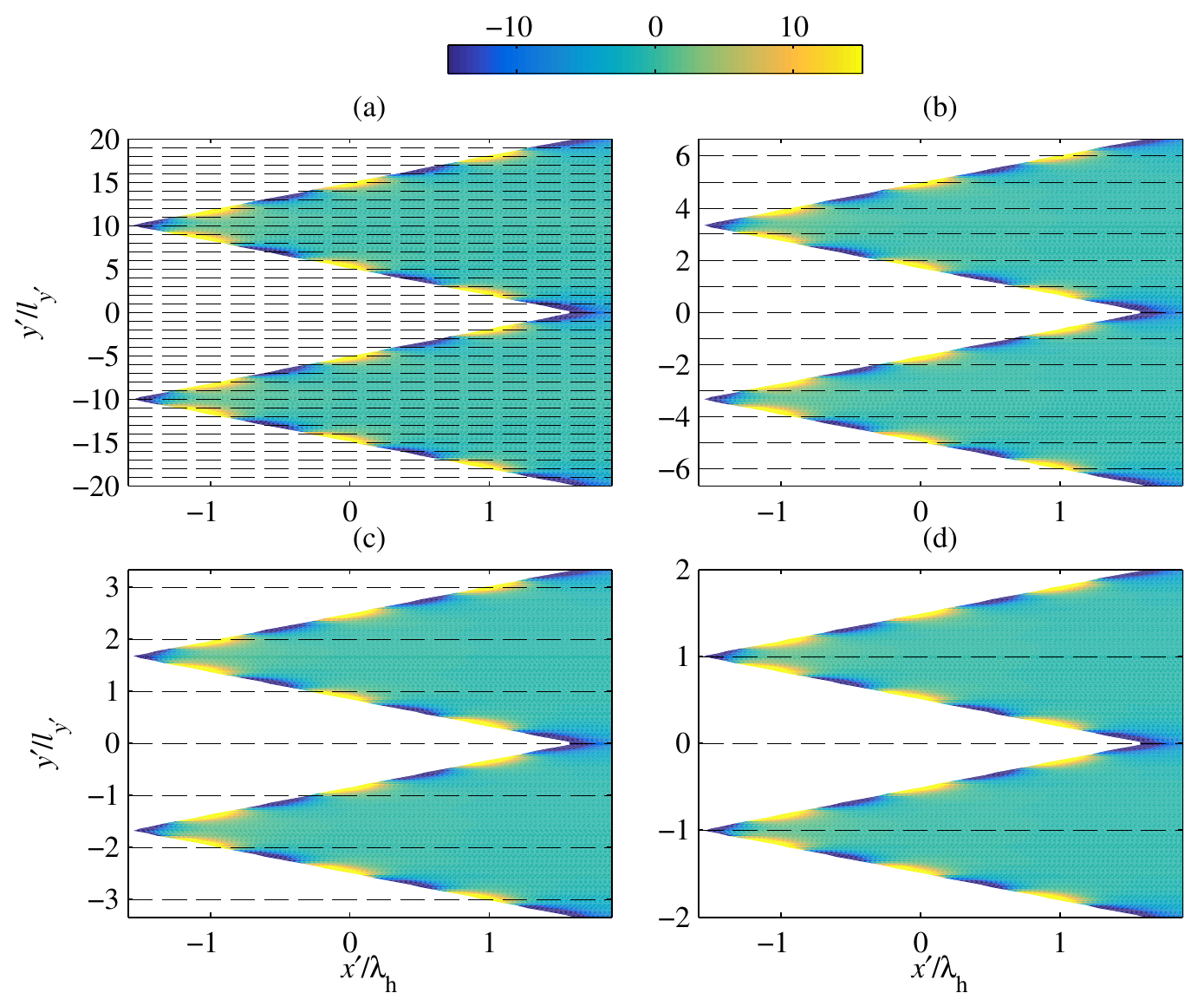}
  \caption{The effects of varying $\omega h_e / U$ on the scattered surface
  pressure at a fixed frequency and $\omega h / U= 10$: a) $\omega h_e / U= 1$;
  b) $\omega h_e / U= 3$; c) $\omega h_e / U = 6$; d) $\omega h_e / U=10$. The
  color shows the normalized scattered pressure on the upper surface of the
  flat plate. The
  vertical axis shows the spanwise coordinate normalized by the spanwise
  correlation length and the horizontal axis shows the streamwise coordinate normalized by the hydrodynamic wavelength.}
  \label{fig:surfacePressureSerratedSharpness}
\end{figure}
high-frequency regimes this assumption should serve as a good approximation, as
demonstrated by both figure~\ref{fig:validation1} and \ref{fig:validation2}.
The importance of the second criterion is also easy to understand. Since the
noise reduction relies on the destructive interference, the scattered pressure
needs to be correlated. $\omega h_e /U \gg 1$ effectively ensures this, hence makes
pressure cancellation due to phase variation possible.

To clearly demonstrate the effect of the $\omega h / U$ criterion, we present
the scattered pressure on the upper surface of the flat plate at fixed $\omega
h_e /U = 7$ in
figure~\ref{fig:surfacePressureSerratedFreq}. To better understand the effect
of sound interference, the $x^\prime$- and $y^\prime$-axis are normalized by
the hydrodynamic wavelength ($\lambda_h = 2\pi U/\omega$) and spanwise correlation length ($l_{y^\prime}$). The distance between the adjacent $y^\prime$-grid lines shown in figure~\ref{fig:surfacePressureSerratedFreq} are used to show the spanwise correlation length $l_{y^\prime}$. When $\omega h / U = 1$, as can be seen from figure~\ref{fig:surfacePressureSerratedFreq}(a), little phase variation is induced along and near the leading edge in the spanwise direction. This suggests little sound reduction occurs due to destructive interference. As $\omega h / U$ increases to $4$, as shown in figure~\ref{fig:surfacePressureSerratedFreq}(b), phase variation of the scattered pressure near the leading edge become noticeable. Further increasing $\omega h / U$ to $10$, we see from figure~\ref{fig:surfacePressureSerratedFreq}(c) that significant sound reduction can be achieved due to the strong phase variation near the leading edge in the spanwise direction. Figure~\ref{fig:surfacePressureSerratedFreq}(d) shows that pressure distribution when $\omega h /U = 20$. It is evident that when $\omega h_e / U $ is fixed, the number of pressure crests and troughs within $l_{y^\prime}$ is fixed at sufficiently large value of $\omega h / U$ (see figures~\ref{fig:surfacePressureSerratedFreq}(b) to (d)). The effects of $\omega h_e / U$ are similarly shown in figure~\ref{fig:surfacePressureSerratedSharpness}, where $\omega h / U$ is fixed at $10$ while $\omega h_e / U $ varies from $1$ to $10$. From figure~\ref{fig:surfacePressureSerratedSharpness}(a) to (d) the scattered pressure patterns are virtually the same. However, the phase variation of the surface pressure within the spanwise correlation length $l_{y^\prime}$ is very different. In figure~\ref{fig:surfacePressureSerratedSharpness}(a) we see that within the length of $l_{y^\prime}$ (between adjacent $y^\prime$-grid lines), the phase variation of pressure is negligible. Since the destructive interference can only occur within $l_{y^\prime}$, little sound reduction can be expected in this case. On the other hand, figure~\ref{fig:surfacePressureSerratedSharpness}(c) and (d) show that a sufficient number of crests and troughs appearing within the length $l_{y^\prime}$, leading to effective noise reduction due to the destructive interference within $l_{y^\prime}$.  Figures~\ref{fig:surfacePressureSerratedFreq} and \ref{fig:surfacePressureSerratedSharpness} clearly show that in order to achieve significant sound reduction, both $\omega h / U \gg 1$ and $\omega h_e / U \gg 1$ have to be satisfied.


\section{Conclusions and future work} \label{sec:Conclusion} 
A new mathematical model is developed in this paper with the aim to predict the
sound radiated from the interaction of an incoming turbulent flow with a flat
plate with serrated leading edge. By making use of the Fourier expansion and
the Schwarzschild techniques, the power spectral density of the far-field sound
is related to the wavenumber spectral density of the incident velocity field.
The model is based on Amiet's approach and is therefore valid even for high
Mach number applications where leading edge noise is a common problem. The
comparisons with experimental data have shown an excellent agreement and this
suggests that the model can capture the essential physics of the noise
generation and reduction mechanisms and can provide accurate prediction of the
noise from serrated leading edges. A thorough parametric study has been carried
out using the new model and the effects of leading edge serration geometry and
incoming turbulent flow characteristics on far-field noise at different Mach
numbers have been studied. It has been found that in order to achieve
significant noise reduction, the serration amplitude $2h$ has to be
sufficiently large compared to the hydrodynamic wavelength in the streamwise
direction. More specifically, the condition of $\omega h / U \gg 1$ needs to be
satisfied. The spanwise correlation length also plays an important role in
achieving effective noise reduction. In order to achieve significant noise
reduction, a second condition of $\omega h_e /U \gg 1$ has to be satisfied,
which ensures that scattered pressure is correlated for a possible destructive
interference to occur.  It has also been shown that leading edge serration can
effectively reduce the far-field noise at even high Mach numbers. However,
larger serrations might be needed especially in the relatively low frequency
regimes. From the far-field noise directivity patterns, it has been observed
that more sound reduction occurs for observer locations near the trailing edge
side of the flat plate and that the noise directivity at high frequencies
changes from the cardioid shape to a tilted dipole-like pattern. The
mathematical model developed in this paper has shown that the destructive sound
interference is the primary noise reduction mechanism, especially in the mid-
to high-frequency regime where leading edge noise is most effectively reduced
using serrations. Further work is needed to address the issue of perfect
coherence for the incoming turbulence in the streamwise direction, which might
not be an accurate assumption at low frequencies.

\section*{Appendix}
    \label{AppendixA}
    \subsection{Second iterated results}
    The 0- and 1-order solutions of $\Phi_t(x, 0)$ are given by
    (\ref{equ:NonscatteringPart}) and \ref{equ:ScatteringPart1}, respectively, in Section~\ref{sec:Formulation}. The 2-order solution is given by 
    $C^{(2)}_{n^\prime}(x)$ as follows
    \begin{equation}
      \begin{aligned}
        C_{n^{\prime}}^{(2)}(x)= & -\Phi_{ia}(1+\i){e}^{{\i}k_1x}
	\sum_{m=-\infty}^\infty\Bigg\{ \beta_{n^{\prime}m} ({\i}k_1)^2 (E^\ast(\mu_{n^{\prime}}x)- E^\ast(\mu_mx))\\
        & + \big(\beta_{n^{\prime}m}{\i}k_1 + \gamma_{n^{\prime}m}
	{\i}(k_1-\mu_m)\big)\sqrt{\frac{\mu_m}{2\pi x}}
	({e}^{-{\i}\mu_{n^{\prime}}x}- {e}^{-{\i}\mu_mx})  \\ 
        & -\frac{\gamma_{n^{\prime}m}}{2}\big(\sqrt{\frac{\mu_m}{2\pi
	x}}\frac{1}{x}({e}^{-{\i}\mu_{n^{\prime}}x}- {e}^{-{\i}\mu_mx})  +
	{\i}(\mu_{n^{\prime}}-\mu_m)\sqrt{\frac{\mu_m}{2\pi
	x}}{e}^{-{\i}\mu_{n^{\prime}}x}\big)\Bigg\},
      \end{aligned}
    \end{equation}
    where
    \begin{equation*}
      \begin{aligned}
        &\beta_{ln} = \sum_{m=-\infty}^{\infty} \big(\alpha_{ln}a_m - B_{lm}/(k_{2l}^2-k_{2n}^2)a_n \big) \alpha_{nm},\\
        &\gamma_{ln} = \sum_{m=-\infty}^{\infty}\big(\alpha_{ln}a_m \sqrt{\mu_m/\mu_n} - B_{lm}/(k_{2l}^2-k_{2n}^2)a_n \big)\alpha_{nm}.\\
      \end{aligned}
    \end{equation*}
    Similarly, $P^{(2)}_{n^\prime}(x)$ can be expressed as
    \begin{IEEEeqnarray*}{rCll}
      P_{n^{\prime}}^{(2)}(x) & = & \rho_0 U\Phi_{ia} & (1+\i)\e^{\i k_1x} \\
      && \sum_{m=-\infty}^\infty\Bigg\{& \beta_{n^{\prime}m} ({\i}k_1)^2
      \frac{1}{\sqrt{2\pi x}} (\sqrt{\mu_{n^\prime}}\e^{-\i\mu_{n^\prime} x} -
      \sqrt{\mu_m} \e^{-\i\mu_m x})\\
      &&& \negmedspace{} -  \big(\beta_{n^{\prime}m}{\i}k_1 +
      \gamma_{n^{\prime}m} {\i}(k_1-\mu_m)\big) \Big[ \i \sqrt{\frac{\mu_m}{2\pi
      x}} (\mu_{n^\prime} \e^{-{\i}\mu_{n^{\prime}}x} - \mu_m \e^{-{\i}\mu_mx}) + \negmedspace{} \\
      &&& \qquad\qquad\qquad\qquad\qquad\qquad \qquad \frac{1}{2}
  \sqrt{\frac{\mu_m}{2\pi x}} \frac{1}{x} ( \e^{-{\i}\mu_{n^{\prime}}x} -
  \e^{-{\i}\mu_mx}) \Big]\\
      &&& \negmedspace{} + \gamma_{n^{\prime}m}
      \Big[\i\frac{1}{2}\sqrt{\frac{\mu_m}{2\pi
      x}}\frac{1}{x}(\mu_{n^\prime}\e^{-{\i}\mu_{n^{\prime}}x} - \mu_m
      \e^{-{\i}\mu_mx}) + \negmedspace{} \\
  &&& \quad\quad\quad\quad \frac{1}{2}\frac{3}{2} \sqrt{\frac{\mu_m}{2\pi
  x}}\frac{1}{x^2}(\e^{-{\i}\mu_{n^{\prime}}x} -  \e^{-{\i}\mu_mx})  + \negmedspace{}\\
      &&&\qquad \quad\ \; \i\frac{(\mu_{n^{\prime}}-\mu_m)}{2}\big(
  \i\mu_{n^\prime} \sqrt{\frac{\mu_m}{2\pi x}}\e^{-{\i}\mu_{n^{\prime}}x} +
  \frac{1}{2}\sqrt{\frac{\mu_m}{2\pi x}}\frac{1}{x} \e^{-{\i}\mu_{n^{\prime}}x} \big) \Big]
    \Bigg\}.\IEEEeqnarraynumspace\IEEEyesnumber
  \end{IEEEeqnarray*}
  \nomenclature[g-beta]{$\beta_{ln}$}{Coefficient in the high-order results}
  \nomenclature[g-gamma]{$\gamma_{ln}$}{Coefficient in the high-order results}
  Finally, $\Theta_{n^\prime}^{(2)}$ can be found as
  \begin{IEEEeqnarray*}{rCll}
    \Theta_{n^\prime}^{(2)} & = & \sum_{m=-\infty}^\infty\Bigg\{&
	\beta_{n^{\prime}m} ({\i}k_1)^2 (\sqrt{\mu_{n^\prime}}S_{n^\prime n^\prime} - \sqrt{\mu_m} S_{n^\prime m})\\
    &&& \negmedspace{} -  \big(\beta_{n^{\prime}m}{\i}k_1 +
    \gamma_{n^{\prime}m} {\i}(k_1-\mu_m)\big) \Big[ \i \sqrt{\mu_m}
	(\mu_{n^\prime} S_{n^\prime n^\prime} - \mu_m S_{n^\prime m} ) +
	\negmedspace{}\\ 
    &&&\quad\quad\quad\quad\quad\quad\quad\quad\quad\quad\quad\quad\quad\quad\sqrt{\mu_m} (T_{n^\prime n^\prime}  -  T_{n^\prime m} ) \Big]\\
    &&& \negmedspace{} + \gamma_{n^{\prime}m} \Big[\i\sqrt{\mu_m}(\mu_{n^\prime}T_{n^\prime n^\prime} - \mu_m T_{n^\prime m} ) +  \sqrt{\mu_m}(V_{n^\prime n^\prime} - V_{n^\prime m})\\
    &&&\qquad \quad\ \; \negmedspace{} +
\i\frac{(\mu_{n^{\prime}}-\mu_m)}{2}\big( \i\mu_{n^\prime} \sqrt{\mu_m}S_{n^\prime n^\prime} + \sqrt{\mu_m} T_{n^\prime n^\prime} \big) \Big]
  \Bigg\}, \IEEEyesnumber
\end{IEEEeqnarray*}
where function $V_{nm}$ is given by
\begin{IEEEeqnarray*}{rCll}
  V_{nm} &=&\sum_{j=0}^1 \Bigg\{& \frac{1}{\sigma_j} \e^{\i \kappa_{ij}
  (\lambda_j + (c-\epsilon_j)/\sigma_j)} \times\negmedspace{}\\
  &&& \quad\quad\quad\Big[\frac{1}{\sqrt{2\pi(c-\epsilon_j)}}\e^{-\i \eta_{Bnmj}
  (c - \epsilon_j)}  - \frac{1}{\sqrt{2\pi(c-\epsilon_{j+1})}}\e^{-\i \eta_{Bnmj} (c - \epsilon_{j+1})}\Big] \\
  &&&\negmedspace{} + \frac{1}{\i\kappa_{nj}}
  \Bigg(\frac{-\i\eta_{Am}}{\sqrt{\eta_{Am}}} \Big[\e^{\i\kappa_{nj}
  \lambda_{j+1}}E^\ast(\eta_{Am}(c-\epsilon_{j+1})) - \e^{\i\kappa_{nj} \lambda_{j}}E^\ast(\eta_{Am}(c-\epsilon_{j}))\Big] + \negmedspace{}\\
  &&&\qquad \qquad \frac{\i\eta_{Bnmj}}{\sqrt{\eta_{Bnmj}}} \e^{\i\kappa_{nj}
  (\lambda_j + (c-\epsilon_j)/\sigma_j)}\times\negmedspace{}\\
  &&& \quad\quad\quad\quad\quad\quad\quad\quad\Big[E^\ast(\eta_{Bnmj}(c-\epsilon_{j+1})) -
  E^\ast(\eta_{Bnmj}(c-\epsilon_{j}))\Big]  \Bigg)\Bigg\}. \\
  \IEEEyesnumber
  \label{equ:V_nm}
\end{IEEEeqnarray*}

\section*{Acknowledgments}
The first author (BL) wishes to gratefully acknowledge the financial
support provided by the Cambridge Commonwealth European and International Trust
and China Scholarship Council. The second author (MA) would like to acknowledge
the financial support of the Royal Academy of Engineering.

\bibliography{referenceslatest}
\bibliographystyle{jfm}
\end{document}